\documentclass[prb,twocolumn,aps,superscriptaddress,floatfix]{revtex4-2}
\usepackage{graphicx} % Required for inserting images
\usepackage{amsbsy,amssymb,amsmath,amsthm,bm,dsfont}
\usepackage[normalem]{ulem}
\usepackage{hyperref}
\hypersetup{
    colorlinks,
    citecolor=black,
    filecolor=black,
    linkcolor=black,
    urlcolor=black
}
\usepackage{placeins}

\theoremstyle{definition}

\begin{document}

\title{Weak-coupling theory of half-metals and other fractional metallic phases in biased doped Bernal bilayer graphene}
\author{D.V. Zhitov}
\affiliation{Abrikosov Center for Theoretical Physics, MIPT, Dolgoprudnyi, Moscow Region 141701, Russia}
\author{A.V. Rozhkov}
\affiliation{Institute for Theoretical and Applied Electrodynamics, Russian Academy of Sciences, 125412 Moscow, Russia}
\author{A.O. Sboychakov}
\affiliation{Institute for Theoretical and Applied Electrodynamics, Russian Academy of Sciences, 125412 Moscow, Russia}
\author{A.L. Rakhmanov}
\affiliation{Institute for Theoretical and Applied Electrodynamics, Russian Academy of Sciences, 125412 Moscow, Russia}

\begin{abstract}
The paper presents a theoretical study of many-body electronic phases in doped and electrically biased Bernal-stacked (AB) bilayer graphene. We develop a variational mean-field theory with no fitted parameters. For the electron-electron interaction, we employ a parameter-free random-phase approximation to model the short-range screened Coulomb repulsion. The remaining long-range Coulomb potential energy is dictated by the geometry of the sample, behaving as a parallel-plate capacitor. We formulate the theory directly in terms of the experimentally controlled displacement field $\frak{D}$ rather than the interaction-renormalized interlayer potential difference. It makes our theory better suited for direct comparison with experiment. The resulting phase diagram is quite rich. It hosts several fractional-metal states connected by first- and second-order transitions. At higher bias and doping, three distinct fractional metallic states emerge. We classify these phases by the number of doped sectors and the symmetry of the order parameters. Our results – obtained without any parameter fitting – reproduce key qualitative and quantitative features of recent experiments. This includes the energy scale associated with the loss of fractional metallic order. At lower bias and doping, the model stabilizes a broad spectrum of fractional-metal phases with more exotic symmetry-breaking patterns. This suggests that further experimental exploration of the latter regime is warranted.
\end{abstract}

\maketitle

\section{Introduction}
Ordered electronic phases in bilayer and multilayer graphene systems have been investigated extensively for more than a decade. This effort spans a broad body of
experimental~\cite{Bao2012, Martin2010, Weitz2010, Mayorov2011,
Freitag2012, Freitag20122053, veligura2012, Velasco2012, freitag2013,
trilayer_quarter2021exper, de_la_barrera_cascade_2022,
Seiler2022, zhou2022isospin}
and
theoretical~\cite{MCCANN2007110, Nandkishore2010, Nandkishore2010b,
vafek_rg2010, vafek_nemat_rg2010, Lemonik2010, Jung2011,
cvetkovic_multi2012, aa_graph_prl2012, haritonov_afm2012,
bilayer_half-metal2013numeric_MF,
baima2018dft_half_met_graphene,
aa_graph_BreyFertig2013, sboychakov2013AA, aa_quarter_met2021}
work; see also the reviews in
Refs.~\onlinecite{bilayer_review2016} and~\onlinecite{Kotov2012RevModPhys}.
A renewed surge of interest in this field has been stimulated, in particular, by
recent experimental observations of fractional metallic phases in doped multilayer
graphene~\cite{trilayer_quarter2021exper, Seiler2022}: isospin-ordered states in which charge carriers populate only a subset of the four spin-valley sectors, rather than distributing themselves symmetrically among all of them. Such states had been anticipated theoretically,
including half-metallic~\cite{bilayer_half-metal2013numeric_MF,
baima2018dft_half_met_graphene}
and quarter-metallic
phases~\cite{aa_quarter_met2021,mayrhofer2025valley}. 

The present work builds upon earlier theoretical
research~\cite{MCCANN2007110, Nandkishore2010, Jung2011,
bilayer_half-metal2013numeric_MF,rozhkov2026bias_cascade}, and in particular extends the framework of Ref.~\onlinecite{rozhkov2026bias_cascade}, which investigated the undoped, insulating AB-BLG, to the doped case. 
Our goal is to formulate an approach that combines analytical transparency with a sufficiently comprehensive account of the principal mechanisms responsible for electronic ordering in AB-BLG. In addition to the standard two-band Hamiltonian with transverse electric bias and exchange electron-electron interaction, the model also incorporates the Hartree
charging energy associated with interlayer polarization. Furthermore, our analysis explicitly retains four independent order parameters, corresponding to different spin and valley sectors. This quartet of order parameters makes it possible to distinguish among several qualitatively different ordered
states in AB-BLG. 

The model is analyzed within a variational mean-field formalism. The resulting self-consistency equations are solved systematically and comprehensively with the help of a symmetry between the four order parameters, and the solutions are compared by their free energy. In the resulting phase diagram we find a cascade of transitions between distinct fractional metallic phases. This leads to a non-monotonic dependence of the single-particle gap on the electric field.

We compare our theoretical predictions with the experimental
results reported in Refs.~\onlinecite{Seiler2022,zhou2022isospin,de_la_barrera_cascade_2022}.
In particular, we construct the phase diagram of the AB-BLG in the plane (doping, bias voltage) without any adjustable parameters and observe that many of its features are consistent with previous experimental results.

The paper is organized as follows. In
Sec.~\ref{sec::model}
we explain our model's structure.
In Sec.~\ref{sec::MF_theory}
the mean field theory is formulated, and the approach to solving self-consistency equations is outlined in Sec. \ref{sec::solving_equations}. The phase diagram is mapped in
Sec.~\ref{sec::results}.
Discussion of the results is in
Sec.~\ref{sec::discussion}.
Conclusions are in
Sec.~\ref{sec::conclusions}.
Several calculations are relegated to Appendices.

\section{The model}
\label{sec::model}
\subsection{Tight-binding model }

Hexagonal lattice of AB bilayer graphene is characterized by unit vectors $\mathbf{a}_{1,2}=a(\sqrt{3},\mp1)/2$, where $a=2.46$\AA . The interlayer distance is $d=3.35$\,\AA. The unit cell contains four atoms, which are identified by the layer number (1 or 2) and sublattice index ($A$ or $B$). Atoms $1B$ and $2A$ sit directly on top of each other, while atoms $1A$ and $2B$  are located above and below the centers of the hexagons of the other layer. The kinetic energy of electrons in this system can be  modeled with a tight-binding Hamiltonian 
\begin{equation}
    H_0=\sum_{\mathbf k\sigma}\psi_{\mathbf k \sigma}^\dagger \left(\mathcal{H}_\mathbf{k}-\mu\right)\psi_{\mathbf k \sigma},
\end{equation}
where $\mu$ is the chemical potential and $\psi^\dagger_{\mathbf k \sigma}$ is a 4-component operator, creating electrons with momentum $\mathbf k$ and spin projection $\sigma$ within sublattices 1A, 1B, 2A, and 2B. Here $\mathcal{H}_{\mathbf k}$ is the  $4\times 4$ matrix 
\begin{equation}
%%%%%%%%%%%%%%%%%%%%%%%%%%%%%%%%%%%%%%%%%%%%%%%%%%
\label{H0}
%%%%%%%%%%%%%%%%%%%%%%%%%%%%%%%%%%%%%%%%%%%%%%%%%%
{\cal H}_{\mathbf{k}}
=
\left(\begin{array}{cccc}
	0&-tf_{\mathbf{k}}&0&t_0\\
	-tf_{\mathbf{k}}^{*}&0&0&0\\
	0&0&0&-tf_{\mathbf{k}}\\
	t_0&0&-tf_{\mathbf{k}}^{*}&0
\end{array}\right),
\end{equation}
with
\begin{equation}
    f_{\mathbf{k}}
=
e^{i\mathbf{k}(\bm{\mathbf{a}_1+\mathbf{a}_2})/3}
\left[1+e^{-i\mathbf{ka}_1}+e^{-i\mathbf{ka}_2}\right].
\end{equation}

Here the amplitudes $t=2.7$\,eV and $t_0=0.4$\, eV describe the in-plane and out-of-plane nearest-neighbor hopping. Because these bonds connect only opposite sublattices, the model is particle–hole symmetric. Further hopping amplitudes are neglected, and with them the splitting of the quadratic band touching at $K$ into four Dirac cones below $\sim 1$ meV. We adopt the present model for its simplicity and tractability; the simplified approach to the band structure is a significant limitation of our work. We return to this point in Sec.~\ref{sec::discussion}.

Exact diagonalization of matrix~\eqref{H0} gives us four energy bands, which we denote with $S=1,\dots,4$. Two of them ($S=2$ and 3) touch parabolically at the $\mathbf K$ and $\mathbf K'$ points of the Brillouin zone while the other two ($S=1$ and 4) are split off with an energy gap $2t_0$. Given that all the experimentally accessible energy scales are usually much smaller than this value, we neglect the outer bands in the following low-energy theory. 
As a result, the kinetic part of the Hamiltonian leads to a theory with 4 species (2 valleys $\times$ 2 bands) of parabolically dispersing spin-1/2 fermions. Denoting their creation operators as $\gamma_{\mathbf{k}S\xi\sigma}$ ($\mathbf{k}$ is the momentum relative to the valley center, $\xi=\pm 1$ is the valley index, $S=2,3$ is the band index, $\sigma=\pm 1$ is the spin index) we can write the low energy Hamiltonian as
\begin{eqnarray}
%%%%%%%%%%%%%%%%%%%%%%%%%%%%%%%%%%%%%%%%%%%%%%%%%%
\label{2band_H}
%%%%%%%%%%%%%%%%%%%%%%%%%%%%%%%%%%%%%%%%%%%%%%%%%%
H_0
=
\sum_{{\bf k} \xi \sigma}
	\left(
		\gamma_{{\bf k} 2 \xi \sigma}^\dag,
		\gamma_{{\bf k} 3 \xi \sigma}^\dag
	\right)
	\left(
		\begin{matrix}
			- \varepsilon_{\bf k}-\mu & 0 \\
			0 & \varepsilon_{\bf k}-\mu \\
		\end{matrix}
	\right)
	\left(
		\begin{matrix}
			\gamma_{{\bf k} 2 \xi \sigma}^{\vphantom{\dag}} \\
			\gamma_{{\bf k} 3 \xi \sigma}^{\vphantom{\dag}} \\
		\end{matrix}
	\right).
\end{eqnarray}
where $\varepsilon_{\mathbf{k}}=\mathbf{k}^2/2m^*$ with $m^*=2t_0/3t^2a^2\approx 0.05m_e$. The momentum summation is cut off at the energy scale $t_0$. 

\subsection{Electron-electron interaction}
In our approach we account for Coulomb interaction, which naturally separates into two physically distinct contributions: a screened, short-range part $\Tilde H_{\rm int}$ and a long-range capacitive term $H^{\rm cap}_{\rm int}$ associated with macroscopic charge imbalance between the layers:
\begin{equation}
H_{\rm int} =
\Tilde H_{\rm int}
+
H^{\rm cap}_{\rm int}
.
\end{equation}

At finite momentum transfer, the Coulomb interaction is efficiently screened by the electron gas. As a result, it becomes effectively short-ranged and primarily mediates scattering between low-energy states within the same valley. Retaining only processes relevant for gap formation, the interaction can be expressed directly in terms of band operators $\gamma_{\mathbf{k} s \xi \sigma}$ as~\cite{rozhkov2026bias_cascade}
\begin{eqnarray}
\label{Hint1}
\Tilde H_{\rm int} =
\frac{-1}{2N_c}
  \sum_{\substack{\mathbf{k}\mathbf{k}'\\ \sigma\sigma'\xi }}
  \Big[
  2\bar \Gamma_{1}\,
  \gamma^{\dagger}_{\mathbf{k}2\xi\sigma}
  \gamma_{\mathbf{k}3\xi\sigma'}
  \gamma^{\dagger}_{\mathbf{k}'3\xi\sigma'}
  \gamma_{\mathbf{k}'2\xi\sigma}\nonumber\\
+
\bar \Gamma_{2}\,
(\gamma^{\dagger}_{\mathbf{k}2\xi\sigma}
\gamma_{\mathbf{k}3\xi\sigma'}
\gamma^{\dagger}_{\mathbf{k}'2\xi\sigma'}
\gamma_{\mathbf{k}'3\xi\sigma}
+
\text{H.c.})
\Big],
\end{eqnarray}
where $N_c$ is the number of unit cells in the sample. 
Here, the two coupling constants $\bar \Gamma_{1,2}$ encode distinct scattering channels.
Coupling $\bar \Gamma_{1}$ describes direct interband scattering, where an electron is transferred between bands $S=2$ and $S=3$.
Meanwhile $\bar \Gamma_{2}$ represents an exchange process, in which the band indices are interchanged.

The coupling constants $\bar \Gamma_{1,2}$ are obtained from an RPA calculation by projection onto the two-band subspace followed by averaging over momentum. 
A more detailed discussion of this approximation can be found in previous work \cite{ab_supercond2023sboychakov}. We are using the values obtained there:
\begin{eqnarray}
%%%%%%%%%%%%%%%%%%%%%%%%%%%%%%%%%%%%%%%%%%%%%%%%%%
\label{Gamma_def}
%%%%%%%%%%%%%%%%%%%%%%%%%%%%%%%%%%%%%%%%%%%%%%%%%%
\bar \Gamma_{1} = 9.37 t,
\quad
\bar \Gamma_{2} = 8.93 t.
\end{eqnarray}

The zero-momentum component of the Coulomb interaction remains unscreened at the macroscopic level and gives rise to an electrostatic energy cost for transferring charge between the layers. This contribution depends only on the total layer polarization 
\begin{eqnarray}
H^{\rm cap}_{\rm int}&&=\frac{{\cal E}_0}{8N_c}(\rho_{10}-\rho_{20})^2
\end{eqnarray}
where
\begin{eqnarray}\label{value_E0}
{\cal E}_0 = \frac{4 \pi e^2 d }{S_0 }
\approx
116\,\text{eV}
\approx
43 t,
\end{eqnarray}
$S_0=\sqrt{3}a^2/2$ is the area of the graphene unit cell, and $\rho_{10}$ and $\rho_{20}$ are operators of the total charge in the top and bottom layer respectively. The charge operators can be expressed in terms of the band operators using the approximate expressions for the wavefunctions:
\begin{equation}
    \rho_{10}-\rho_{20}=-\sum_{{\bf k} \xi \sigma}
	\left(\gamma_{{\bf k} 2 \xi \sigma}^\dag
	\gamma_{{\bf k} 3 \xi \sigma}^{\vphantom{\dag}}
	+
	\gamma_{{\bf k} 3 \xi \sigma}^\dag
	\gamma_{{\bf k} 2 \xi \sigma}^{\vphantom{\dag}}\right)
\end{equation}
This interaction penalizes net imbalance between the layers, acting as a capacitive energy proportional to the square of the interlayer charge polarization.

\subsection{External field}
Two external gates allow one to independently control the electron density in the bilayer and the potential difference between the layers. The latter effect is described with the displacement field $\frak{D}$. The contribution to the Hamiltonian is
\begin{eqnarray}
    H_\Phi = -\frac{e\Phi}{2} \sum_{{\bf k} \xi \sigma}
	\left(\gamma_{{\bf k} 2 \xi \sigma}^\dag
	\gamma_{{\bf k} 3 \xi \sigma}^{\vphantom{\dag}}
	+
	\gamma_{{\bf k} 3 \xi \sigma}^\dag
	\gamma_{{\bf k} 2 \xi \sigma}^{\vphantom{\dag}}\right),
\end{eqnarray}
where $\Phi$ is $\frak{D}d$. Some care should be taken when defining this quantity due to the fact that layer polarization is itself a source of the electric field besides the control gates. A careful discussion of the macroscopic electrostatics of the system can be found in Appendix \ref{appendix:electrostatics}.

The full Hamiltonian is
\begin{widetext}
\begin{eqnarray}
\label{eq::full_hamiltonian}
    H&=&H_0
-\frac{1}{2N_c}
  \sum_{\substack{\mathbf{k}\mathbf{k}'\\ \sigma\sigma'\xi }}
  \Big[
  2\bar \Gamma_{1}\,
  \gamma^{\dagger}_{\mathbf{k}2\xi\sigma}
  \gamma_{\mathbf{k}3\xi\sigma'}
  \gamma^{\dagger}_{\mathbf{k}'3\xi\sigma'}
  \gamma_{\mathbf{k}'2\xi\sigma}
+
\bar \Gamma_{2}\,
(\gamma^{\dagger}_{\mathbf{k}2\xi\sigma}
\gamma_{\mathbf{k}3\xi\sigma'}
\gamma^{\dagger}_{\mathbf{k}'2\xi\sigma'}
\gamma_{\mathbf{k}'3\xi\sigma}
+
\text{H.c.})
\Big]
\\
&-&%\frac{{\cal E}_0}{8N_c}(\rho_{10}+\rho_{20})^2+
\frac{{\cal E}_0}{8N_c}
\left[
\sum_{\mathbf{k}\xi\sigma}
\left(\gamma^{\dagger}_{\mathbf{k}2\xi\sigma}\gamma_{\mathbf{k}3\xi\sigma}
+
\gamma^{\dagger}_{\mathbf{k}3\xi\sigma}\gamma_{\mathbf{k}2\xi\sigma}
\right)\right]^2
-\frac{e\Phi}{2} \sum_{{\bf k} \xi \sigma}
	\left(\gamma_{{\bf k} 2 \xi \sigma}^\dag
	\gamma_{{\bf k} 3 \xi \sigma}^{\vphantom{\dag}}
	+
	\gamma_{{\bf k} 3 \xi \sigma}^\dag
	\gamma_{{\bf k} 2 \xi \sigma}^{\vphantom{\dag}}\right).\nonumber
\end{eqnarray}
\end{widetext}
It is studied in the following sections.

\section{Mean field theory}
\label{sec::MF_theory}

We restrict our considerations to mean field approach developed in Ref.~\onlinecite{rozhkov2026bias_cascade} with the order parameters associated with interlayer polarization in each valley. In a given valley, the order parameter is $2\times 2$ matrix defined as the expectation value of the operator
\begin{eqnarray}
\Xi_{{\bf k} \xi}^{\sigma \sigma'}
=
\gamma^{\dag}_{\mathbf{k}3{\xi \sigma}}
\gamma^{\phantom{\dag}}_{\mathbf{k} 2\xi \sigma'}.
\end{eqnarray}
We use a variational Hamiltonian of the form
\begin{eqnarray}
%%%%%%%%%%%%%%%%%%%%%%%%%%%%%%%%%%%%%%%%%%%%%%%%%%
\label{eq::MF_Hamiltonian_def}
%%%%%%%%%%%%%%%%%%%%%%%%%%%%%%%%%%%%%%%%%%%%%%%%%%
H_{\rm MF}
=
H_0
- \sum_{{\bf k} \xi}
	{\rm Tr}\!  \left(
		\hat{\Delta}_{ \xi}^{\vphantom{\dag}}
		\hat\Xi_{{\bf k} \xi}^\dag
		+
		\hat\Xi_{{\bf k} \xi}^{\vphantom{\dag}}
		\hat{\Delta}_{ \xi}^\dag
	\right),
\end{eqnarray}
where the gap parameters $\hat\Delta_\xi$ are $2\times2$ \textit{c}-number matrices and they capture the effects of interaction, doping, and the bias voltage. We assume that $\hat\Delta_\xi$ is Hermitian since previous works suggest that non-Hermitian $\hat\Delta_\xi$ are associated with inter-plane currents and are less favorable energetically. \cite{ rozhkov2023aa_su4, rozhkov2025ab_su4}

As usual in mean-field theory we derive two complementary self-consistency equations. The first one concerns the expectation values with respect to the variational wave function, which is the ground state of $H_{\rm MF}$ for a given $\hat\Delta_\xi$. A straightforward calculation with the Hamiltonian~\eqref{eq::MF_Hamiltonian_def} at zero temperature gives (assuming without loss of generality $\mu>0$)
\begin{equation}
    \langle \hat\Xi_{\mathbf k\xi}\rangle =\frac{1}{2}\sum_{i=1}^2{(v_{\xi i} v_{\xi i}^\dagger)}\frac{D_{\xi i}}{\sqrt{\varepsilon_{\mathbf k}^2+D_{\xi i}^2}}\Theta\left(\sqrt{\varepsilon_{\mathbf k}^2+D_{\xi i}^2}-\mu\right),
\end{equation}
where $D_{i\xi}$ and $v_{i\xi}$ are the eigenvalues and normalized eigenvectors of the matrix $\hat\Delta_\xi$, and $\Theta(x)$ is the Heaviside step function. The details of derivation can be found in Appendix~\ref{appendix:self_consistency_derivation}.

The second equation follows from minimization of the variational energy with respect to $\hat\Delta_{\xi}$:

\begin{eqnarray}\label{Delta_Eq}
\hat{\Delta}_{ \xi}^{\vphantom{\dag}}
=
\frac{1}{N_c}
\sum_{\mathbf{k}'}\left(
	\bar \Gamma_{1}
	\langle \hat\Xi_{{\bf k}' \xi}^{\vphantom{\dag}} \rangle
	+
	\bar \Gamma_{2} \langle \hat\Xi_{{\bf k}' \xi}^\dag \rangle\right)
-
\\
\nonumber 
	\frac{{\cal E}_0}{4} \hat{\mathbb{I}}_2
	\sum_{\xi'}
		{\rm Tr}
		\langle
			\hat\Xi_{{\bf k}' \xi'}^{\vphantom{\dag}}
			+
			\hat\Xi_{{\bf k}' \xi'}^\dag
		\rangle+\frac{e\Phi}{2}\hat{\mathbb{I}}_2.
\end{eqnarray}

We can reduce two self-consistency equations to a single one by eliminating $\hat\Xi_{\mathbf k\xi}$ and taking into account the Hermiticity of the matrices $\hat\Delta_\xi$ (see Appendix \ref{appendix:self_consistency_derivation} for derivation). As a result, we obtain
\begin{widetext}
\begin{eqnarray}
\label{eq:self_consistency_dimensionful}
    \frac{1}{2N_c}\sum_{\mathbf k}\frac{D_m}{\sqrt{\varepsilon_{\mathbf k}^2+D_m^2}}\Theta\!\left({\sqrt{\varepsilon_{\mathbf k}^2+D_m^2}-\mu}\right) = 
    \frac{D_m}{\bar \Gamma_1+\bar \Gamma_2}
    -\frac{\mathcal{E}_0}{2(\bar \Gamma_1\!+\!\bar \Gamma_2)(2\mathcal{E}_0\!-\!\bar \Gamma_1\!-\!\bar \Gamma_2)}\sum_{m'}D_{m'}
    +\frac{e\Phi}{2(2\mathcal{E}_0\!-\!\bar \Gamma_1\!-\!\bar \Gamma_2)},
\end{eqnarray}
\end{widetext}
where multi-index $m = (i,\xi)$ enumerates four fermionic sectors.

For further consideration, it is useful to introduce the following parameters:
\begin{eqnarray}
\gamma&=&(\bar{\Gamma}_1+\bar{\Gamma}_2)\nu_0,\,\,\,
\Delta_0=2t_0 e^{-2/\gamma},\,\,\,
\delta_0=\frac{\Delta_0}{2t_0},\nonumber\\
\bar{\varepsilon}&=&{\cal E}_0\nu_0,\,\,\,
\lambda_{\rm FE}=\bar{\varepsilon}-\frac12 \gamma,\,\,\,
\Lambda = \frac{\bar{\varepsilon}}{2\lambda_{\rm FE}\gamma},\nonumber
\\
x_m &=& \frac{D_m}{\Delta_0},\,\,\,
M = \frac{\mu}{\Delta_0},\,\,\,
V = \frac{1}{\lambda_{\rm FE}}\frac{e \Phi/2} {\Delta_0},\label{parameters}
\end{eqnarray}
where $\nu_0=t_0/2\sqrt{3}\pi t^2$ is the density of states per unit cell of undoped AB-BLG. In the third line in Eq.~\eqref{parameters}, the quantity $x_m$ is the dimensionless order parameter in sector $m$, while $M$ and $V$ are the dimensionless chemical potential and dimensionless bias voltage, respectively. Both $M$ and $V$ can change in experiment by application of the voltage to the gates located above and below bilayer. The parameter $\Delta_0$ in Eq.~\eqref{parameters} is the ground state order parameter of the undoped and unbiased bilayer~\cite{rozhkov2026bias_cascade}. We calculate all the variables in first two lines of Eq.~\eqref{parameters} without using any adjustable parameters as  
\begin{eqnarray}
\gamma&=&0.249,\,\,\,\Delta_0=0.26\,\rm{meV}=3\,\rm{K},\,\,\,
\delta_0=3.2\times10^{-4},\nonumber\\
\bar{\varepsilon}&=&0.585,\,\,\,
\lambda_{\rm FE}=0.460,\,\,\,
\Lambda=2.55.\nonumber
\end{eqnarray}
We see that in the case of the AB-BLG the weak-coupling approach is relevant since the dimensionless coupling parameter $\gamma$ is small. We also obtain the value of the order parameter $\Delta_0$ close to the experimentally measured~\cite{Seiler2022,Geisenhof2022} value. 

We perform summation over momenta in the left-hand side of Eq.~\eqref{eq:self_consistency_dimensionful} 
following a standard procedure with constant density of states $\nu_0$ and the energy cutoff $t_0$. As a result, in the limit $\delta_0\ll1$, we obtain
\begin{eqnarray}\label{eq::self_consistency_equation}
   &&x_m\ln\left(\frac{1}{|x_m|}\right)-x_m\ln\left(\frac{M+\sqrt{M^2-x_m^2}}{|x_m|}\right)\times\nonumber\\
   &&\Theta(|M|-|x_m|)=V-\Lambda(x_1+x_2+x_3+x_4).
\end{eqnarray}
This equation has to be solved with respect to $x_m$ for given $M$ and $V$. The set of $x_m$ fully specifies the expectation value $\langle \hat \Xi_{\mathbf k\xi}\rangle $ up to unitary transformations in the spin space. 

The number of extra electrons per unit area is
\begin{equation}
    n_e=\frac{\nu_0\Delta_0}{S_0}\sum_m\sqrt{M^2-x_m^2}\Theta(M-|x_m|).
\end{equation}
For analytical convenience, we also define its dimensionless counterpart $n\equiv n_{e}S_0/(\nu_0\Delta_0)$.

\section{Solutions of the self-consistency equation}
\label{sec::solving_equations}

The self-consistency equation (\ref{eq::self_consistency_equation}) is a system of four equations for four variables $x_m$. 
In this system the dependence of the solution on a particular $x_m$ is fully controlled by its left-hand side, while the right-hand side depends on the sum of $x_m$. Thus, Eq.~\eqref{eq::self_consistency_equation} can be presented in the form
\begin{equation}
    C(x_m) = V-\Lambda\sum_n x_n,
\end{equation}
where 
\begin{equation}\label{eq:C(x)}
    C(x)\equiv x\ln\left(\frac{1}{|x|}\right)-x\ln\left(\frac{M+\sqrt{M^2-x^2}}{|x|}\right)\!\Theta(M-|x|).
\end{equation}
Such a specific symmetry of the self-consistency equations dictates a specific strategy to solve them.
The details of the numerical procedure can be found in Appendix~\ref{appendix:numerical_procedure}.

The self-consistency equations have several distinct solutions $\{x_m\}$ at some values of $M$ and $V$, 
which indicates the existence of unstable and metastable phases. We compare corresponding free energies to identify the ground state. 
This procedure should be performed in the canonical (fixed particle number) rather than grand canonical (fixed chemical potential) ensemble,
since the former condition better matches the AB-BLG experiment. 
A more detailed discussion of these issues can be found in the Appendix \ref{appendix:electrostatics}.

The Helmholtz free energy is defined as
\begin{eqnarray}
    F = \Omega+\mu\langle N\rangle,
\end{eqnarray}
where $\Omega=\langle H\rangle$ is the grand canonical free energy and the Hamiltonian $H$, Eq.~(\ref{eq::full_hamiltonian}), includes the chemical potential.
We calculate $\Omega=\langle H\rangle$ in the basis of the mean-field variational wave function using Wick's theorem
\begin{eqnarray}\label{omega_1}
    \frac{\Omega}{\nu_0\Delta_0^2N_c}
    &=&
    \frac{\langle H_0\rangle}{\nu_0\Delta_0^2N_c}+\sum_m (x_m-\lambda_{\rm FE}V)L_m\nonumber\\
    &-&\frac{\gamma}{4}\sum_m L_m^2
    +\frac{\bar\varepsilon}{8}\left(\sum_m L_m\right)^2,
    \end{eqnarray}
where 
\begin{eqnarray}
    \!\!\! L_m = \frac{1}{\nu_0\Delta_0N_c}\sum_{\mathbf k}\!\frac{D_m}{\sqrt{\varepsilon
    _{\mathbf k}^2+D_m^2}}\Theta\!\left(\sqrt{\varepsilon_{\mathbf k}^2+D_m^2}-\mu\right)\!.
\end{eqnarray}
With the help of the self-consistency equation~(\ref{eq::self_consistency_equation}) we obtain
\begin{equation}\label{Lm}
    L_m=V+\frac{2}{\gamma}x_m-\Lambda\sum_n x_n.
\end{equation}
Performing summation in Eq.~\eqref{2band_H} and, in the standard way, moving from summation over momentum to integration over energy we derive for $\langle H_0\rangle$ 
\begin{eqnarray}\nonumber
    \langle H_0\rangle=
    \nu_0\Delta_0^2N_c \sum_m\!\left\{
   \!-\!\int\limits_{\varepsilon_0}^{1/2\delta_0}\sqrt{x_m^2+\varepsilon^2}\,d\varepsilon\!-\!M n
    \right\},
\end{eqnarray}
where $\varepsilon_0=\sqrt{\max(0, M^2-x_m^2)}$. Substituting the expression for $L_m$ and $\langle H_0\rangle$ in Eq.~\eqref{omega_1} and performing integration over $\varepsilon$ we obtain
\begin{widetext}
\begin{eqnarray}
    \frac{F}{\nu_0\Delta_0^2N_c}=\sum_m\left[-\frac12 \varepsilon\sqrt{x_m^2+\varepsilon^2}+\frac12x_m^2 \ln(\sqrt{x_m^2+\varepsilon^2}-\varepsilon)
    \right]_{\varepsilon=\varepsilon_0}^{\varepsilon=1/2\delta_0}\nonumber\\
    +\frac{1}{\gamma}\sum_m x_m^2 -\frac{\Lambda}{2}\left(\sum_m x_m\right)^2+V\sum_m x_m - 2\lambda_{\rm FE}V^2.
\end{eqnarray}
\end{widetext}
This is the final expression for the free energy that will be used to construct the phase diagram of the model.

\section{Results}
\label{sec::results}

As doping increases at fixed bias field, our model predicts that the system passes through a sequence of distinct ordered phases in which progressively more spin-valley sectors become populated with charge carriers, until full symmetry among all four sectors is restored at large doping. We describe the resulting phase diagram in detail below, before comparing it with experiment in Sec.~\ref{sec::discussion}.

In order to compare the predictions of our model with recent experimental measurements of fractional-metallic AB-BLG~\cite{zhou2022isospin, de_la_barrera_cascade_2022, Seiler2022}, we compute and plot the phase diagram in the plane $(n_e,\frak{D})$ for $n_e\in [-0.8, 0]\times 10^{12}\rm{cm}^{-2}$, $\frak{D}\in [0,1.1]\, \rm{V/nm}$. We consider the case of hole doping ($n_e<0$) since it corresponds to the experimental data. Our model is particle-hole symmetric, so accounting for hole doping involves trivial changes in signs of $\mu$ and $n_e$ in our formulas. We also note that here we do not consider the insulating phases with $n=0$. They are qualitatively different from fractional metals and their properties have already been discussed in Ref.~\onlinecite{rozhkov2026bias_cascade}.

\begin{figure}
    %\centering
    \includegraphics[width=1.0\linewidth]{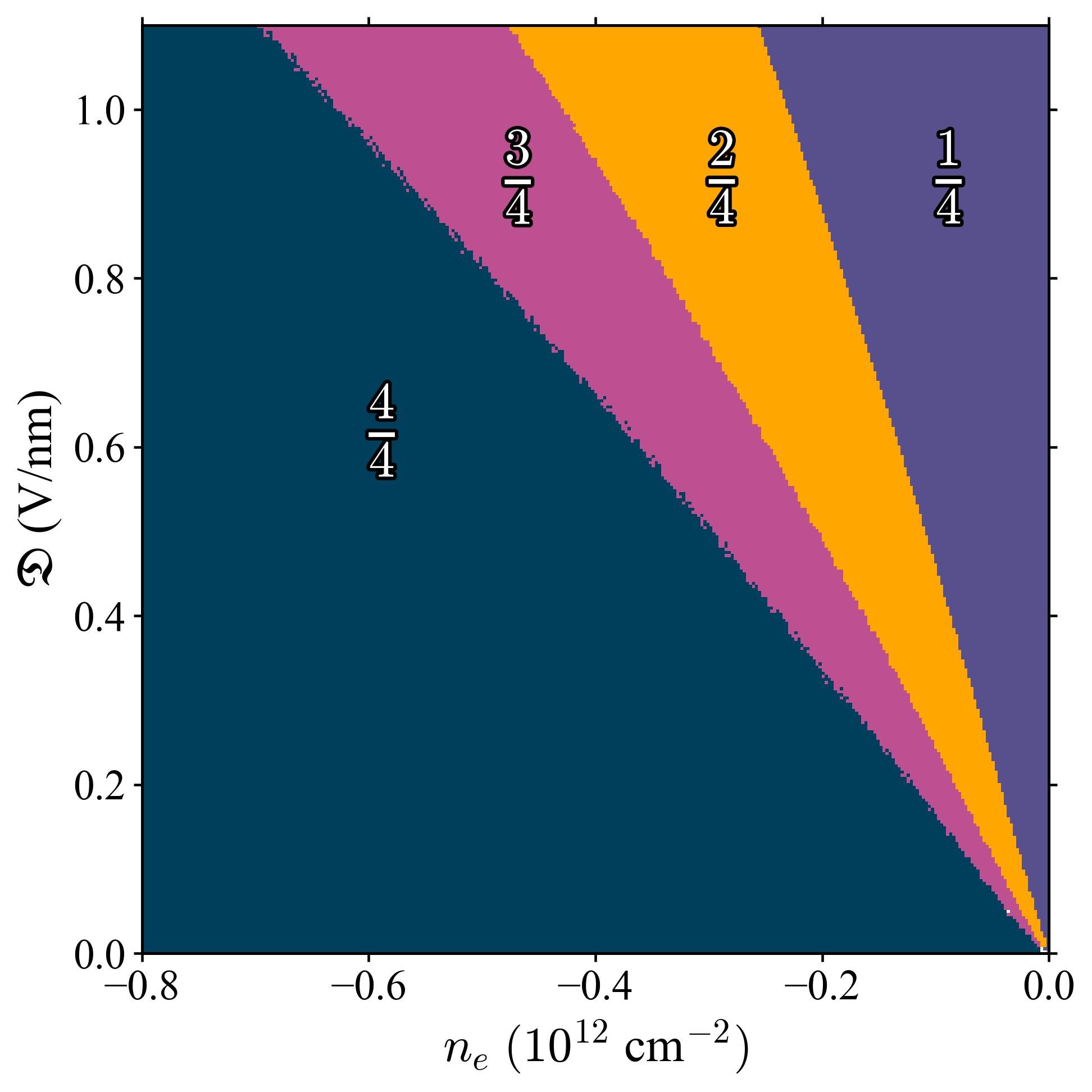}
    \caption{Phase diagram of AB-stacked bilayer graphene within the experimentally accessible range: the labeled phases are (from right to left) quarter-metal, half-metal, three quarter-metal, and metal. The phases are separated by first-order transitions.}
    \label{fig:phase_diagram_full}
\end{figure}

The phase diagram obtained numerically is shown in Figure~\ref{fig:phase_diagram_full}. The phase boundaries correspond to the singular behavior of the gap in the electron spectrum caused by first-order phase transitions. The phase boundaries exhibit a characteristic downward bend at higher values of $|n_e|$ and $\frak{D}$.

We group the phases into four categories: metal, three-quarters-metal, half-metals, quarter-metals. The classification is based on the number of spin-valley sectors doped in the given phase. In each of these phases, the gap parameters $x_m$ have no more than two distinct values. Therefore, the Fermi surfaces in all doped sectors are identical.

\emph{Quarter-metal.} The quarter-metal state is observed at the lowest values of doping. This state is fully polarized in spin-valley degrees of freedom, and thus can be called an isospin ferromagnet in the language of Ref.~\onlinecite{zhou2022isospin}. It spontaneously breaks the discrete valley symmetry and the continuous spin-rotation symmetry.

\emph{Half-metal.} In the considered mean-field approach, the self-consistency equations are degenerate with respect to permutation of the spin and valley indices. Thus,  the correspondence between the spin and valley degrees of freedom and the sectors of our model is not unique. Among the states we consider, this issue is only relevant for the half-metal. It could either be spin-polarized, e.g. by having Fermi surfaces at $(K,\uparrow)$ and $(K', \uparrow)$ , valley-polarized – $(K,\uparrow)$ and $(K,\downarrow)$, or spin-valley-polarized – $(K,\uparrow)$ and $(K',\downarrow)$. To find the ground state unambiguously, we need to expand our theory taking into account neglected factors.

\emph{Three quarters-metal.} In the phase observed in our numerical phase diagram, three of the four sectors have identical doping, while the fourth one is not doped. To our knowledge, such a phase has not been previously seen in experiments; we return to this discrepancy, and a likely explanation for it, in Sec.~\ref{sec::discussion}.

\emph{Metal.} In this phase all the sectors are equivalent, and all the symmetries are restored.

\begin{figure}
    \centering
    \includegraphics[width=1.0\linewidth]{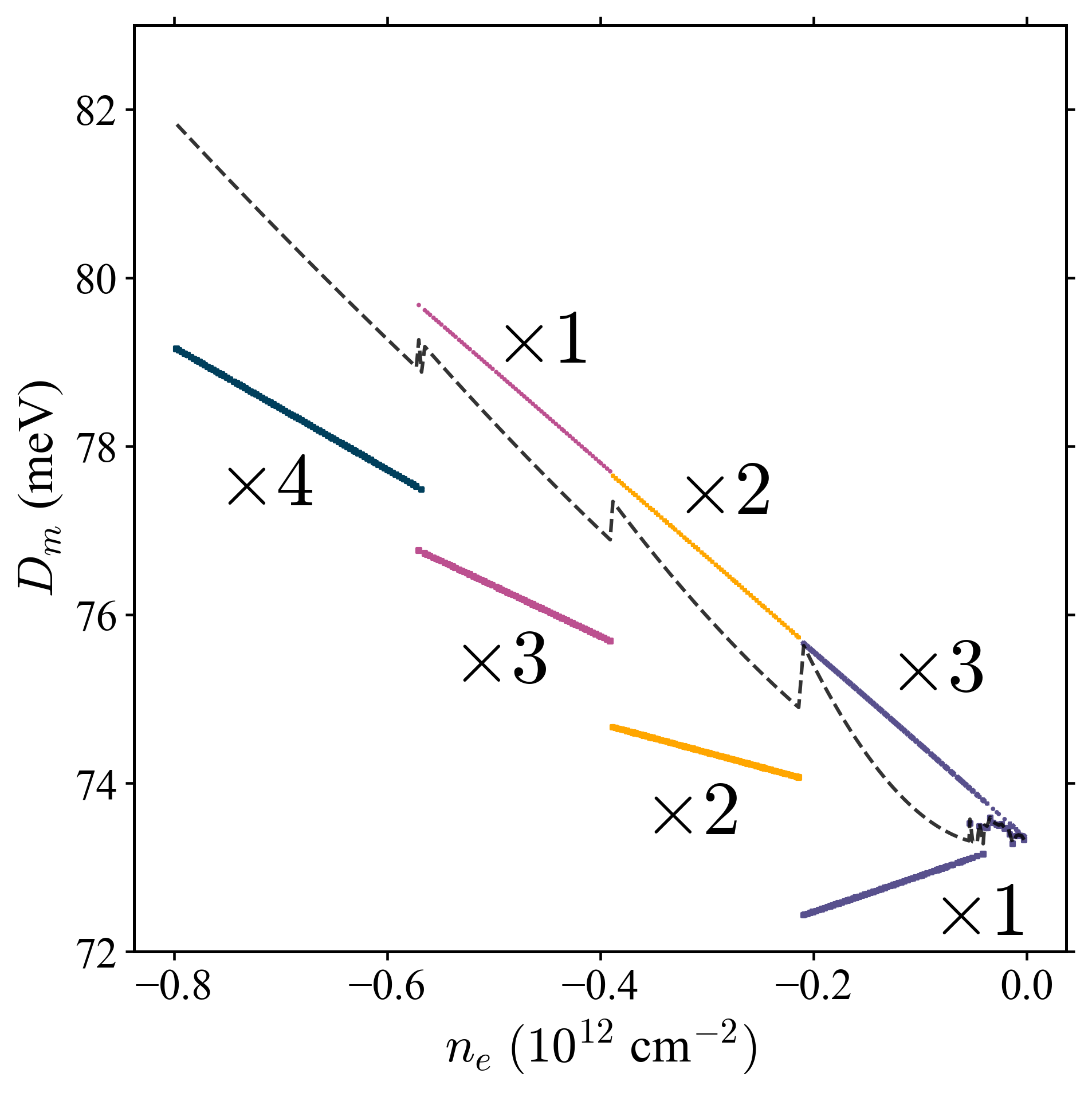}
    \caption{Gap parameters of the four sectors as a function of $n_e$ for $\frak{D}=0.92$ V/nm. Dashed line denotes the chemical potential. The degeneracies associated with each value of the gap parameter are labeled. The degeneracy of the gaps below the chemical potential  corresponds to the number of doped sectors of the fractional metallic phases.} 
    \label{fig:cross_section}
\end{figure}

The variation of the four gap parameters $x_m$ with doping at fixed applied voltage is shown in Fig.~\ref{fig:cross_section}. One can clearly see the transitions between the different phases, involving jumps of the chemical potential (dashed line) and changes of the number of doped sectors. 
It is worth noting that the values of the gap parameters follow $\mu$ rather closely in the sense that $|D_m-\mu|\ll \mu$. This is because the gap is controlled primarily by the transverse field (screened by layer polarization) rather than doping. Moreover, the band $E_m(\mathbf k)=\sqrt{\varepsilon_{\mathbf{k}}^2+D_m^2}$ is relatively shallow for the typical values of the parameters, so small relative differences between $D_m$ and $\mu$ are sufficient to achieve the necessary doping.

\begin{figure}
    \centering
    \includegraphics[width=1.0\linewidth]{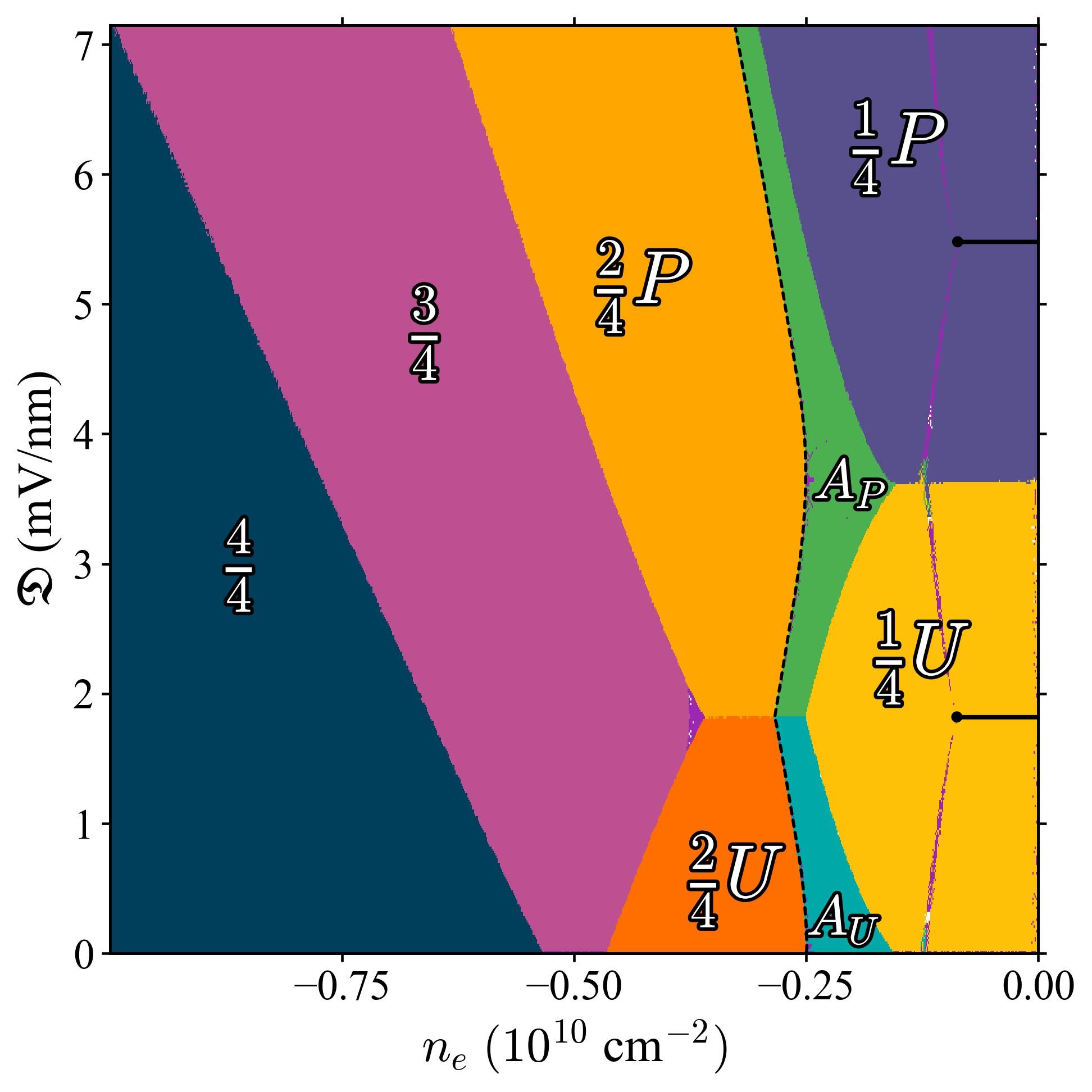}
    \caption{Phase diagram at low $n_e$ and $\frak{D}$. Four half-metallic ($\frac24 P,\frac24 U,A_P,A_U$) and two quarter-metallic ($\frac14 P,\frac14 U$) phases are identified. The continuous symmetry-breaking transition between the phases $\frac24 P$, $A_P$ and $\frac24 U$, $A_U$ are denoted with dashed lines. All other boundaries correspond to first-order transitions.
    Within the quarter-metallic phases there are horizontal first-order transition lines, separating the phases with identical symmetry but different values of the gap parameters and the chemical potential. Several numerical artifacts are present; these are discussed in the main text.}
    \label{fig:phase_diagram_small}
\end{figure}

We also observe a more rich and intricate phase diagram at low doping and low electric field, Fig.~\ref{fig:phase_diagram_small}. In dimensionless units these additional features arise at $n\sim 1$ and $V\sim\Lambda$ (cf. $n\sim 200$, $V\sim 1500$ above). These phases can also be subdivided into four categories according to the number of doped sectors.
We find four types of a half-metal and two 
types of a quarter-metal. We classify these phases according to (a) signs of the gap parameters $x_m$ in the undoped sectors, (b) symmetry or asymmetry of $x_m$ in the doped sectors. The number of doped sectors corresponds to the fractional metallicity described above.  For half-metals and quarter-metals criterion (a) allows one to separate the observed phases further. We denote the corresponding possibilities with letters P (polarized) and U (unpolarized). In P-phases the gap parameters $x_m$ in the undoped sectors are all positive and equal to each other (there are two of those in a half-metal and three in a quarter metal). In U-phases one of them is negative while the remaining ones (one for half-metal and two for quarter-metal) are still positive. Finally, phases $\frac{2}{4}P$ and $\frac{2}{4}U$ are continuously connected to phases $A_P$ and $A_U$ where the symmetry of the doped sectors gets spontaneously broken: in the latter phases the gap parameters $x_m$ in the two doped sectors have different values, while in all the other phases the doped sectors are always identical. The competition between the polarised and unpolarised phases can be seen as a competition between the capacitive energy and the effect of the external field.

The low doping – low field phase diagram contains several artifacts related to limitations of the numerical procedure. This includes the almost vertical curves around $n_e\approx 10^9 \,{\rm cm}^{-2}$ in Fig. \ref{fig:phase_diagram_small} which touch the critical points. The parts of the phase diagram separated by these curves do not exhibit any sharp changes. Another region of interest is the triangular feature at $n_e\approx 4\times 10^9\,{\rm cm}^{-2}$, $\frak D\approx 2 \, {\rm mV/nm}$. This region results from competition among several phases that are nearly degenerate in energy, making its detailed structure particularly sensitive to modifications of the model. We therefore consider the precise structure of this feature insufficiently robust to warrant a more detailed investigation in the present work.

\section{Discussion}
%%%%%%%%%%%%%%%%%%%%%%%%%%%%%%%%%%%%%%%%%%%%%%%%%% 
\label{sec::discussion}
%%%%%%%%%%%%%%%%%%%%%%%%%%%%%%%%%%%%%%%%%%%%%%%%%% 

\subsection{Key features of our approach}

The framework presented in this work combines three essential ingredients.
The first is the
short-range screened Coulomb repulsion, described by the interaction
constants $\bar\Gamma_1$ and $\bar\Gamma_2$ obtained from an RPA
calculation without adjustable parameters. The second is the long-range
capacitive energy $H^{\rm cap}_{\rm int}$ associated with interlayer
charge imbalance. The latter is crucial for a meaningful comparison with
experiment: it accounts for the screening of the applied displacement
field by the induced layer polarization, allowing us to formulate the
theory directly in terms of the experimentally controlled quantity
$\frak{D}$ rather than an effective interlayer potential difference. 
The third one is electron-hole (excitonic) ordering between the two bands. It occurs via a BCS-like mechanism and has the typical logarithmic divergence that leads to ordering even in the weakly-coupled regime.
All
ingredients are treated within a controlled mean-field
approximation: the dimensionless coupling constant $\gamma = 0.249$
[Eq.~\eqref{parameters}] is small, placing the system in the regime where
the mean-field treatment represents the leading order in a systematic
expansion. The combination of these ingredients leads to the fractional metallic phases.

The mechanism described above is quite dissimilar to the Stoner
criterion commonly employed in the analysis of isospin ordering in
multilayer graphene~\cite{Raines2024,mayrhofer2025valley}.
We regard the Stoner criterion and the present approach as fundamentally
different in their status. The Stoner criterion, by construction, can
always account for an observed polarization provided the interaction is
chosen sufficiently strong. To illustrate this point, consider
Ref.~\onlinecite{mayrhofer2025valley}
where the coupling constants 
$U_1 \sim U_2 \sim 240$ eV\,\AA$^2$
have been fixed by matching an experimental feature. We note that the density of states of the
relevant band in AB-BLG falls between
$\sim 0.025$~eV$^{-1}$\AA$^{-2}$
and
$\sim 0.2$~eV$^{-1}$\AA$^{-2}$,
see the plots in Fig.~8 of
Ref.~\onlinecite{mayrhofer2025valley}.
The dimensionless coupling constants
$\nu U_{1,2}$
are therefore never below $\sim 6$ and may be as large as $\sim 48$. At
such coupling strengths, not only does the Stoner criterion lack systematic
justification, but the very notion of well-defined quasiparticles is not
guaranteed. Unless a specialized approach tailored to this specific
strong-coupling problem is devised, little can be said about the system's
properties with confidence. This illustrates the general difficulty: within
the Stoner framework, agreement with experiment may often be achieved by
selecting a sufficiently large interaction, making it difficult to assess
the significance of such
agreement~\cite{Holzmann2020}.

\subsection{Comparison with experiment}

Our results partially agree with the recent experiments, see
Refs.~\onlinecite{de_la_barrera_cascade_2022}, \onlinecite{Seiler2022},
and \onlinecite{zhou2022isospin}. First, we see fractional metallic phases
on the phase diagram. The shape of the phase boundaries, with bending down
at higher values of $|n_e|$ and $\frak{D}$, qualitatively agrees with the
experimental observations. Moreover, our predictions for the magnitude of
the energy gap are qualitatively consistent with the experiments. In
particular, the estimated ordering scale
$\Delta_0\approx 3$~K
compares well with the temperature
($\sim 2-3$~K)
at which the phase boundaries
smear~\cite{zhou2022isospin}.
As a result, the calculated phase boundaries on the plane
$(n_e,\frak{D})$
are quantitatively consistent with measurements. We stress that this
agreement is obtained without any adjustable parameters: every quantity
entering the theory is calculated either from the tight-binding model
itself or from an independent RPA treatment of the screened Coulomb
interaction.

Our model, however, does not capture all the phases identified in the
experiments~\cite{zhou2022isospin}. In particular, we do not find the
so-called
$\rm{Sym}_{12}$
phases or the partially-isospin-polarized
$\rm{PIP}_2$
phase. Furthermore, we obtain a three-quarter-metal phase in
Fig.~\ref{fig:phase_diagram_full}
that has not been observed in the experiments. A likely origin of these
discrepancies is the simplifications we introduced when constructing our
Hamiltonian: isotropic dispersion, electron-hole symmetry, no intervalley
interaction, and, perhaps, disorder.

\subsection{Outlook}

Several natural extensions of the present work suggest themselves. The
most immediate is the incorporation of trigonal warping terms in the band
structure, which would bring the model closer to the full tight-binding
dispersion and may allow access to the partially-isospin-polarized phases
observed experimentally. A second extension is the treatment of
intervalley interaction channels, which we have neglected but which may
modify the competition between various fractional-metallic states. Finite-temperature effects could be studied within
the same mean-field framework to map out the thermal evolution of the
phase boundaries and compare with the experimentally observed smearing of
transitions.

The mechanism of ordering in
isospin space identified here has broader implications that go beyond  AB-BLG. Similar fractional metallic phases have
been observed~\cite{trilayer_quarter2021exper, Seiler2022} in rhombohedral trilayer graphene and
other graphene multilayers. The framework
developed here can be adapted to these systems to clarify the universality
and system-specificity of the phase diagrams.

Finally, we note that the low-field, low-doping region of our phase
diagram, with its multiple exotic fractional-metal phases, remains largely
unexplored experimentally. We hope that the present theoretical
predictions will motivate further measurements in this regime.

\section{Conclusions}
\label{sec::conclusions}

We have developed a simple mean-field description of the electronic properties of doped AB stacked bilayer graphene (AB-BLG) in the presence of a transverse electric field. In addition to short-range screened Coulomb interactions, the model incorporates the electrostatic energy arising due to interlayer polarization. The theory further includes four independent order parameters, each corresponding to a particular combination of valley and spin quantum numbers, while current-carrying order parameters are excluded from the outset. We derive and solve the resulting set of four self-consistency equations.

Although the ordered phases differ substantially in their physical nature, many of them are found to be degenerate. These phases can nevertheless be classified according to their spin and valley structure. The applied electric field drives transitions between distinct ordered states of the system. The single-particle gap exhibits a non-monotonic dependence on the bias, and, at fixed bias, the gaps associated with different spin and/or valley sectors may differ from one another.

The calculated phase diagram indicates that upon an increase of doping, the system passes through a series of fractional metallic states in which the spin-valley sectors get doped one by one. At high bias the gap parameters of the doped sectors are all equal to each other and the same is true for the undoped sectors.
The framework proposed here is a promising approach to the fractional metallic physics of AB-BLG systems that is likely to provide a reliable description upon more detailed treatment of the band structure effects.

\appendix
\section{Electrostatics of dual-gated bilayer}
\label{appendix:electrostatics}

The experimental setup used for the measurements can be conceptualized as follows. There are two conducting plates, one above and one below the bilayer. The voltage can be applied independently to the plates and bilayer. We assume that the net charge density on the plates and the bilayer is zero:
\begin{eqnarray}
    \sigma_{\rm t}+\sigma_{\rm b}+\sigma_{\rm bl}=0,
\end{eqnarray}
where the subscripts $\rm t$, $\rm b$, and $\rm bl$ stand for the top and bottom plates, and the bilayer, respectively. Note that $\sigma_{\rm bl}=e(\rho_{10}+\rho_{20})/S$ where $e<0$ is the charge of an electron and $S=N_cS_0$ is the sample area. It is convenient to put zero voltage at the bilayer, $V_{\rm bl}=0$.
The considered system includes three regions: the spaces between the bilayer and the top and bottom plates, and the space between the two atomic layers of the bilayer. The displacement fields in the first two regions can be determined according to the Gauss's law as
\begin{equation}
    \frak{D}_{\rm t} = 4\pi\sigma_{\rm t},\quad \frak{D}_{\rm b}=-4\pi\sigma_{\rm b},
\end{equation}
while the electric field between the layers $E_{\rm in}$ is related to them via
\begin{equation}
    E_{\rm in}=\frak{D}_{\rm t}+4\pi e\frac{\rho_{10}}{S}=\frak{D}_{\rm b}-4\pi e\frac{\rho_{20}}{S}.
\end{equation}

The free energy of the system is 
\begin{eqnarray}\label{eq:electric_free_energy}
    W = \frac{S}{8
    \pi\varepsilon}\left(\frak{D}_{\rm t}^2d_{\rm t}+\frak{D}_{\rm b}^2d_{\rm b}\right) + \frac{S}{8\pi}E_{\rm in}^2d\nonumber\\
    -\sigma_{\rm t}  S V_{\rm t}-\sigma_{\rm b} S V_{\rm b}+\langle H_0+\tilde H_{\rm int}\rangle,
\end{eqnarray}
where $d_{\rm t,b}$ are the distances from the bilayer to the top and bottom gates, $\varepsilon$ is the relative permittivity of the material between the gates and the bilayer ($\varepsilon=1$ for suspended samples and $\varepsilon\approx 3\,–\,4$ for hBN encapsulated samples). The first two terms are the electric field energy, the following two are the work of the voltage sources, and the last term corresponds to the energy of the electrons, including the kinetic energy and the screened interaction.

The minimization of the free energy $W$ can be performed conveniently in two steps. At the first step we minimize $W$ with respect to electronic degrees of freedom while keeping $\frak{D}_{t,b}$ constant (note that they are fixed by these gate charges). This variation allows for different values of bilayer polarization and, thus, of the field $E_{\rm in}$. The procedure yields an effective free energy of the bilayer as a function of $\frak{D}_{\rm t},\frak{D}_{\rm b}$. At the second step the free energy $W$ can be minimized with respect to $\frak{D}_{\rm t}$ and $\frak{D}_{\rm b}$ in order to relate the latter to the gate voltages.

One should note that upon the variation at fixed $\frak{D}_{\rm t}$ and $\frak{D}_{\rm b}$ the total electron density on the bilayer is fixed since $4\pi e(\rho_{10}+\rho_{20})/S=\frak{D}_{\rm b}-\frak{D}_{\rm t}$. This is the reason for choosing to minimize the Helmholtz free energy and fixed $n_e$ instead of  the grand canonical free energy at fixed $\mu$.

The field between the layers can be rewritten in a more symmetric form
\begin{equation}
    E_{\rm in}={\frak{D}}+2\pi e\frac{\rho_{10}-\rho_{20}}{S},
\end{equation}
where
\begin{equation}
    {\frak{D}}\equiv\frac{\frak{D}_{\rm t}+\frak{D}_{\rm b}}{2}.
\end{equation}
This leads to an expression for the free energy in which the capacitive and potential terms are clearly separated: 
\begin{eqnarray}\label{eq:electric_free_energy_expanded}
    W &=&
    \frac{S}{8\pi\varepsilon}\left(\frak{D}_{\rm t}^2d_{\rm t}+\frak{D}_{\rm b}^2d_{\rm b}\right)
    +
    \frac{S}{8\pi}{\frak{D}}^2d
    +
    \frac{e{\frak{D}}d}{2} (\rho_{10}-\rho_{20})
    \nonumber\\
    &&{}+
    \frac{\pi e^2 d}{2S}(\rho_{10}-\rho_{20})^2
    +
    \langle H_0+\tilde H_{\rm int}\rangle    
    \nonumber\\
    &&{}-
    \sigma_{\rm t}  S V_{\rm t}
    -
    \sigma_{\rm b} S V_{\rm b}
    ,
\end{eqnarray}
and this provides a direct link to the Hamiltonian \eqref{eq::full_hamiltonian} under identification $\Phi={\frak{D}}d$.

Our definition of $\frak{D}$ as the arithmetic mean of the top and bottom displacement fields has a clear theoretical advantage of allowing one to express the polarization effects purely in terms of the difference $\rho_{10}-\rho_{20}$. Moreover, it is consistent with the definitions used in some experimental works.

For example, de la Barrera et al. define the field $\frak D$ as $(C_{\rm t}V_{\rm t}-C_{\rm b}V_{\rm b})/2\varepsilon_0$ \cite{de_la_barrera_cascade_2022}. In our notation it becomes
\begin{equation}
    \frak{D}_{\rm exp}\equiv \frac{\varepsilon}{2}\left(\frac{V_{\rm t}}{d_{\rm t}}-\frac{V_{\rm b}}{d_{\rm b}}\right).
\end{equation}
We relate the voltages and the fields in the second step of the minimization, varying the free energy $W$ with respect to $\frak{D}_{\rm t,b}$ assuming corresponding adjustment of the electronic degrees of freedom. The latter implies
\begin{eqnarray}
    \left(\frac{\partial\langle H\rangle}{\partial (\frak{D}_{\rm t}+\frak{D}_{\rm b})}\right)_{\frak{D}_{\rm t}-\frak{D}_{\rm b}}
    &=&
    \frac{ed}{4}(\rho_{10}-\rho_{20}),
    \\
    \left(\frac{\partial\langle H\rangle}{\partial (\frak{D}_{\rm t}-\frak{D}_{\rm b})}\right)_{\frak{D}_{\rm t}+\frak{D}_{\rm b}}
    &=&
    \frac{\mu S}{4\pi e},
\end{eqnarray}
which finally yield
\begin{eqnarray}
  \frac{1}{\varepsilon}\frak{D}_{\rm t}d_{\rm t}+\frac{\frak{D}d}{2}+\frac{\pi ed}{S}(\rho_{10}-\rho_{20})-\frac{\mu}{e} &=&V_{\rm t}, \\
  \frac{1}{\varepsilon}\frak{D}_{\rm b}d_{\rm b}+\frac{\frak{D}d}{2}+\frac{\pi ed}{S}(\rho_{10}-\rho_{20})+\frac{\mu}{e} &=&-V_{\rm b}.
\end{eqnarray}
These two equations relate the electric fields and the voltages. The first three terms are geometric: they are equal to the line integrals of the electric field from the mid-plane of the bilayer to the gates with a field $\frak{D}_{t(b)}/\varepsilon$ along the distance $d_{t(b)}$ and a field $E_{\rm in}$ along a distance $d/2$. The final term is the difference between electrochemical and electric potential in graphene bilayer.

Using these expressions we find
\begin{equation}
    \frak D_{\rm exp}=\frak{D}
    +
    \varepsilon E_{\rm in}\frac{d}{2}\left(\frac{1}{d_{\rm t}}+\frac{1}{d_{\rm b}}\right)
    -
    \frac{\varepsilon\mu}{e}\left(\frac{1}{d_{\rm t}}-\frac{1}{d_{\rm b}}\right).
\end{equation}

As one can see, $\frak D_{\rm exp}=\frak{D}$ to leading order. The other two terms reflect the dependence of the $V=0$ equipotential surface on the polarization and doping of the bilayer. Fortunately, the second term can be neglected due to bilayer thickness being much smaller that the distance to the gates, while the last one is identically zero for the symmetric gates, used in most of the experiments.

\section{Derivation the self-consistency equations}
\label{appendix:self_consistency_derivation}
\subsection{First self-consistency equation}

The Hamiltonian $H_{\rm MF}$, Eq.~\eqref{eq::MF_Hamiltonian_def}, can be rewritten explicitly as
\begin{widetext}
\begin{equation}
    H_{\rm MF}=\sum_{\mathbf k,\xi}
    \begin{pmatrix}
        \gamma_{\mathbf k 2\xi\uparrow}^\dagger&
        \gamma_{\mathbf k 2\xi\downarrow}^\dagger&
        \gamma_{\mathbf k 3\xi\uparrow}^\dagger&
        \gamma_{\mathbf k 3\xi\downarrow}^\dagger
    \end{pmatrix}
    \begin{pmatrix}
            (-\varepsilon_{\mathbf k}-\mu)\hat{\mathds{1}}_{2} & -
            \hat\Delta_\xi\\
            -\hat\Delta_\xi^\dagger & (\varepsilon_{\mathbf k}-\mu)\hat{\mathds{1}}_{2}
    \end{pmatrix}
    \begin{pmatrix}
        \gamma_{\mathbf k 2\xi\uparrow}\\
        \gamma_{\mathbf k 2\xi\downarrow}\\
        \gamma_{\mathbf k 3\xi\uparrow}\\
        \gamma_{\mathbf k 3\xi\downarrow}\\
    \end{pmatrix}.
\end{equation}
\end{widetext}
We diagonalize the Hamiltonian using a Bogoliubov transformation
\begin{equation}
    \begin{pmatrix}
        \gamma_{\mathbf k 2\xi\uparrow}\\
        \gamma_{\mathbf k 2\xi\downarrow}\\
        \gamma_{\mathbf k 3\xi\uparrow}\\
        \gamma_{\mathbf k 3\xi\downarrow}\\
    \end{pmatrix}=U_{\mathbf k\xi}\begin{pmatrix}
        b_{\mathbf k \xi 1}\\
        b_{\mathbf k \xi 2}\\
        b_{\mathbf k \xi 3}\\
        b_{\mathbf k \xi 4}
    \end{pmatrix},
\end{equation}
where $U_{\mathbf k\xi}$ is a $4\times 4$ unitary matrix. The corresponding eigenvalues are 
\begin{equation}
    E_{i}=\pm\sqrt{\varepsilon^2+D_{1,2}^2}-\mu, 
\end{equation}
where $D_{1,2}$ are the two eigenvalues of the matrix $\hat\Delta$ (assumed to be Hermitian) and the subscript $i=1,\dots,4$ represents all possible choices of the $\pm$ sign and $D_{1,2}$ (we omit subscripts $(\mathbf k,\xi)$ for brevity). The columns of $U$ form a set of four orthonormal vectors of the Hamiltonian. 

Now we can proceed to the calculation of $\langle\hat\Xi\rangle$. Since $\gamma$-operators are linear combinations of the $b$-operators, expectation values of products of the former can be easily expressed in terms of expectation values of products of the latter. At zero temperature, the states with $E<0$ are filled, while those with $E>0$ are empty. This means that
\begin{equation}
    \langle \gamma_i^\dagger \gamma_j\rangle = \sum_{n,m}U^*_{im}U_{jn}\langle b_m^\dagger b_n\rangle = \sum_n U^*_{in}U_{jn}\Theta(-E_n).
\end{equation}
Direct calculations give the elements of the transformation matrix $U_{\mathbf k\xi}$. As a result we obtain
\begin{equation}
    \langle \Xi^{\sigma\sigma'}\rangle =\langle \gamma_{3\sigma}^\dagger \gamma_{2\sigma'}\rangle =-\frac{1}{2}\sum_{i=1}^4\frac{D_i}{E_i+\mu}v_{\xi i}^{*\sigma}v_{\xi i}^{\sigma'}\Theta(-E_i),
\end{equation}
where $v_{\xi i}^{\sigma}$ are the components of the $i$-th normalized eigenvector of $\hat\Delta_\xi$.
We can consider further only the case $\mu>0$ without loss of generality since our model obeys the particle-hole symmetry. In so doing, we get 
\begin{equation}
\label{eq:first_self_consistency}
    \langle \hat\Xi_{\mathbf k\xi}\rangle =+\frac{1}{2}\sum_{i=1}^2{(v_{\xi i} v_{\xi i}^\dagger)}\frac{D_{\xi i}}{\sqrt{\varepsilon_{\mathbf k}^2+D_{\xi i}^2}}\Theta\left(\sqrt{\varepsilon_{\mathbf k}^2+D_{\xi i}^2}-\mu\right),
\end{equation}
and the expression for the trace is
\begin{eqnarray}
%\label{eq:first_self_consistency}
        {\rm Tr} \langle \hat\Xi_{\mathbf k\xi}\rangle =\frac{1}{2}\sum_{i=1}^2\frac{D_{\xi i}}{\sqrt{\varepsilon_{\mathbf k}^2+D_{\xi i}^2}}\Theta\left(\sqrt{\varepsilon_{\mathbf k}^2+D_{\xi i}^2}-\mu\right).
\end{eqnarray}

\subsection{Second self-consistency equation}

We minimize the energy $\langle H\rangle$.
Using Eq.~\eqref{eq::MF_Hamiltonian_def} we write
$$\langle H_0 \rangle=\left\langle H_{\rm MF}+\sum_{{\bf k} \xi}
	{\rm Tr}\!  \left(
		\hat{\Delta}_{ \xi}^{\vphantom{\dag}}
		\hat\Xi_{{\bf k} \xi}^\dag
		+
		\hat\Xi_{{\bf k} \xi}^{\vphantom{\dag}}
		\hat{\Delta}_{ \xi}^\dag
	\right)\right\rangle$$
and obtain
\begin{eqnarray}
%%%%%%%%%%%%%%%%%%%%%%%%%%%%%%%%%%%%%%%%%%%%%%%%%%
\label{H0_diff}
%%%%%%%%%%%%%%%%%%%%%%%%%%%%%%%%%%%%%%%%%%%%%%%%%%
\partial \langle H_0 \rangle 
=
\partial \langle H_{\rm MF}\rangle
+
\sum_{\bf k}
	\langle
		\hat\Xi_{{\bf k} \xi}^*
	\rangle
+
\\
\nonumber
\sum_{\bf k}
{\rm Tr}\! \left(
	\partial
		\langle \hat\Xi_{{\bf k} \xi}^{\dag}  \rangle
	\hat{\Delta}_{ \xi}^{\vphantom{\dag}}
	+
	\hat{\Delta}_{ \xi}^\dag
	\partial
		\langle \hat\Xi_{{\bf k} \xi}^{\vphantom{\dag}}  \rangle
\right)\!,
\end{eqnarray}
where $\partial X$ means $\partial X/\partial \left[ \hat \Delta_\xi \right]_{\sigma \sigma'}$.
The first two terms in the right-hand side of Eq.~(\ref{H0_diff}) cancel according to the Hellmann–Feynman theorem. Therefore, we have
\begin{eqnarray}\label{eq::difer_H0}
\partial \langle H_0 \rangle
=
\sum_{\bf k}
	{\rm Tr}\!
	\left(
		\hat{\Delta}_{ \xi}^{\vphantom{\dag}}
		\partial \langle \hat\Xi_{{\bf k} \xi}^{\dag}  \rangle
		+
		\hat{\Delta}_{ \xi}^\dag
		\partial \langle
			\hat\Xi_{{\bf k} \xi}^{\vphantom{\dag}}
		\rangle
	\right),
\end{eqnarray}

Next we differentiate
$\langle H-H_0 \rangle = \langle \Tilde H_{\rm int} + H^{\rm cap}_{\rm int}+H_\Phi\rangle $, and get
\begin{widetext}
\begin{eqnarray}
%%%%%%%%%%%%%%%%%%%%%%%%%%%%%%%%%%%%%%%%%%%%%%%%%%
\label{interaction_diff}
%%%%%%%%%%%%%%%%%%%%%%%%%%%%%%%%%%%%%%%%%%%%%%%%%%
\partial \langle H-H_0 \rangle
=
\frac{{\cal E}_0}{4N_c}
\sum_{{\bf k}' \xi' }
\left(	
		{\rm Tr} \langle
				\hat\Xi_{{\bf k}' \xi'}^{\vphantom{\dag}}
			\rangle
			+
		{\rm Tr} \langle \hat\Xi_{{\bf k}' \xi'}^\dag \rangle
\right)
\sum_{\bf k}
\left(	
		{\rm Tr}\,
		\partial \langle
				\hat\Xi_{{\bf k} \xi}^{\vphantom{\dag}}
			\rangle
		+
		{\rm Tr}
		\partial \langle
				\hat\Xi_{{\bf k} \xi}^\dag
			\rangle
\right)
-
\\
\nonumber
\frac{1}{N_c}
\sum_{\mathbf{kk}'}\left[
	\bar \Gamma_{1} {\rm Tr}\!
	\left(
		\langle \hat\Xi_{{\bf k} \xi}^\dag \rangle
		\partial \langle
				\hat\Xi_{{\bf k}' \xi}^{\vphantom{\dag}}
			\rangle
		+
		\langle \hat\Xi_{{\bf k}' \xi}^{\vphantom{\dag}} \rangle
		\partial \langle \hat\Xi_{{\bf k} \xi}^\dag \rangle
	\right)
	+
	\bar \Gamma_{2} {\rm Tr}\!
	\left(
		\langle \hat\Xi_{{\bf k} \xi}^\dag \rangle
		\partial \langle \hat\Xi_{{\bf k}' \xi}^\dag \rangle
		+
		\langle \hat\Xi_{{\bf k} \xi}^{\vphantom{\dag}} \rangle
		\partial
		\langle \hat\Xi_{{\bf k}' \xi}^{\vphantom{\dag}} \rangle
	\right)\right]\\
    -\frac{e\Phi}{2}\sum_{\bf k}
	{\rm Tr}\!
	\left(
		\partial \langle \hat\Xi_{{\bf k} \xi}^{\dag}  \rangle
		+
		\partial \langle
			\hat\Xi_{{\bf k} \xi}^{\vphantom{\dag}}
		\rangle
	\right)\nonumber.
\end{eqnarray}
Combining the expressions for
$\partial \langle H_0 \rangle$
and
$\partial \langle H-H_0 \rangle$,
we derive
\begin{eqnarray}
\partial \langle H\rangle 
=
\sum_{\bf k}
	{\rm Tr} \left\{
		\partial \langle \hat\Xi_{{\bf k} \xi}^{\dag}  \rangle
	\left[
		%\left(
        \hat{\Delta}_{ \xi}^{\vphantom{\dag}}-\frac{e\Phi}{2}\hat{\mathbb{I}}_2
        %\right)
		-
		\frac{1}{N_c}
		\sum_{\mathbf{k}'}
        \left(
			\bar \Gamma_{1}
			\langle
				\hat\Xi_{{\bf k}' \xi}^{\vphantom{\dag}}
			\rangle
			+
			\bar \Gamma_{2}
			\langle \hat\Xi_{{\bf k}' \xi}^\dag \rangle
        \right.\right.\right.
        \nonumber\\
        \left.\left.\left.{}
		+\frac{{\cal E}_0}{4}
		\left(
			{\rm Tr}
			\langle
				\hat\Xi_{{\bf k}' \xi'}^{\vphantom{\dag}}
			\rangle
			+
			{\rm Tr}
			\langle \hat\Xi_{{\bf k}' \xi'}^\dag \rangle
		\right) \hat{\mathbb{I}}_2
        \right)
	\right]
	\right\}
+ {\rm C.c.}
\end{eqnarray}
Thus, $\partial \langle H\rangle=0$ if the expression in the square brackets vanishes, and the extrema of the variational energy correspond to
\begin{eqnarray}\label{OP_Tr_cap_Gamma_0}
\hat{\Delta}_{ \xi}^{\vphantom{\dag}}
=
\frac{1}{N_c}
\sum_{\mathbf{k}'}\left(
	\bar \Gamma_{1}
	\langle \hat\Xi_{{\bf k}' \xi}^{\vphantom{\dag}} \rangle
	+
	\bar \Gamma_{2} \langle \hat\Xi_{{\bf k}' \xi}^\dag \rangle
-
	\frac{{\cal E}_0}{4} \hat{\mathbb{I}}_2
	\sum_{\xi'}
		{\rm Tr}
		\langle
			\hat\Xi_{{\bf k}' \xi'}^{\vphantom{\dag}}
			+
			\hat\Xi_{{\bf k}' \xi'}^\dag
		\rangle
        \right)
    +
    \frac{e\Phi}{2}\hat{\mathbb{I}}_2,
\end{eqnarray}
which coincides with Eq.~\eqref{Delta_Eq} in the main text.

\subsection{Elimination of $\langle\Xi\rangle$ in the self-consistency equations}

Equations~\eqref{eq:first_self_consistency} and \eqref{OP_Tr_cap_Gamma_0} can be considered as a self-consistency equations in our mean field approximation. Yet it is more convenient to invert them, expressing
$\sum_{\bf k} \langle \hat\Xi_{{\bf k} \xi}^{\vphantom{\dag}} \rangle$
and
$\sum_{\bf k} \langle \hat\Xi_{{\bf k} \xi}^\dag \rangle$
as functions of
$\hat \Delta^{\vphantom{\dag}}_\xi$
and
$\hat \Delta^\dag_\xi$.
In this case, only one term that is nonlinear in $D_{m}$ remains in the final self-consistency equation.

As a first step, we define
\begin{eqnarray}
%%%%%%%%%%%%%%%%%%%%%%%%%%%%%%%%%%%%%%%%%%%%%%%%%%
\label{D_defin}
%%%%%%%%%%%%%%%%%%%%%%%%%%%%%%%%%%%%%%%%%%%%%%%%%%
\hat{\cal D}_\xi
=
\hat\Delta_\xi
+
\frac{{\cal E}_0}{4}
\frac{1}{N_c}
\sum_{{\bf k}'\xi}
	{\rm Tr}\,
	\langle
		\hat\Xi_{{\bf k}' \xi'}^{\vphantom{\dag}}
		+
		\hat\Xi_{{\bf k}' \xi'}^\dag
	\rangle
\hat{\mathbb{I}}_2
- \frac{e \Phi}{2} \hat{\mathbb{I}}_2.
\end{eqnarray}
The matrices
$\hat{\cal D}_\xi^{\vphantom{\dag}}$ and $\hat{\cal D}_\xi^\dag$
satisfy the relations that follow from Eq.~\eqref{OP_Tr_cap_Gamma_0}
\begin{eqnarray}
%%%%%%%%%%%%%%%%%%%%%%%%%%%%%%%%%%%%%%%%%%%%%%%%%%
\label{D_VS_Xi}
%%%%%%%%%%%%%%%%%%%%%%%%%%%%%%%%%%%%%%%%%%%%%%%%%%
\hat{\cal D}_\xi^{\vphantom{\dag}}
=
\frac{1}{N_c}
\sum_{{\bf k}'}\left(
	\bar\Gamma_{1}
	\langle \hat\Xi_{{\bf k}' \xi}^{\vphantom{\dag}} \rangle
	+
	\bar\Gamma_{2} \langle \hat\Xi_{{\bf k}' \xi}^\dag \rangle\right),
\quad
\hat{\cal D}_\xi^\dag
=
\frac{1}{N_c}
\sum_{{\bf k}'}\left(
	\bar\Gamma_{1} \langle \hat\Xi_{{\bf k}' \xi}^\dag \rangle
	+
	\bar\Gamma_{2}
	\langle \hat\Xi_{{\bf k}' \xi}^{\vphantom{\dag}}  \rangle\right).
\end{eqnarray}
We obtain after simple algebra from Eqs.~\eqref{D_VS_Xi}
\begin{eqnarray}\label{SumSum}
\frac{1}{N_c}
\sum_{{\bf k}'}
	\langle \hat\Xi_{{\bf k}' \xi}^{\vphantom{\dag}} \rangle
=
\frac{1}{\bar \Gamma_1^2 - \bar \Gamma_2^2}
\left(
	\bar \Gamma_1 \hat{\cal D}_\xi^{\vphantom{\dag}}
	-
	\bar \Gamma_2 \hat{\cal D}_\xi^\dag
\right),
\quad
\frac{1}{N_c}
\sum_{{\bf k}'}
	\langle \hat\Xi_{{\bf k}' \xi}^\dag \rangle
=
\frac{1}{\bar \Gamma_1^2 - \bar \Gamma_2^2}
\left(
	\bar \Gamma_1 \hat{\cal D}_\xi^\dag
	-
	\bar \Gamma_2 \hat{\cal D}_\xi^{\vphantom{\dag}}
\right).
\end{eqnarray}
Substituting definitions~(\ref{D_defin}) in Eqs.~\eqref{SumSum}, we derive
\begin{eqnarray}
%%%%%%%%%%%%%%%%%%%%%%%%%%%%%%%%%%%%%%%%%%%%%%%%%%
\label{Xi_VS_Delta2}
%%%%%%%%%%%%%%%%%%%%%%%%%%%%%%%%%%%%%%%%%%%%%%%%%%
\frac{(\bar \Gamma_1 + \bar \Gamma_2)}{N_c}
\sum_{{\bf k}'}
	\langle \hat\Xi_{{\bf k}' \xi}^{\vphantom{\dag}} \rangle
=
\frac{1}{\bar \Gamma_1 - \bar \Gamma_2}
\left(
	\bar \Gamma_1 \hat \Delta_\xi^{\vphantom{\dag}}
	-
	\bar \Gamma_2 \hat\Delta_\xi^\dag
\right)
+
\frac{{\cal E}_0}{4}
\frac{1}{N_c}
\sum_{{\bf k}'\xi}
	{\rm Tr}\,
	\langle
		\hat\Xi_{{\bf k}' \xi'}^{\vphantom{\dag}}
		+
		\hat\Xi_{{\bf k}' \xi'}^\dag
	\rangle
\hat{\mathbb{I}}_2
-
\frac{e \Phi}{2}
\hat{\mathbb{I}}_2,
\end{eqnarray}
Then, we eliminate the trace of $\langle \hat\Xi_{\mathbf{k}\xi} \rangle$ in the right-hand side of the latter formula. 
To express this trace in terms of the order parameter matrix, we take trace of both sides of Eq.~(\ref{Xi_VS_Delta2})
\begin{eqnarray}
%%%%%%%%%%%%%%%%%%%%%%%%%%%%%%%%%%%%%%%%%%%%%%%%%%
\label{Xi_VS_Delta_Tr1}
%%%%%%%%%%%%%%%%%%%%%%%%%%%%%%%%%%%%%%%%%%%%%%%%%%
\frac{(\bar \Gamma_1 + \bar \Gamma_2)}{N_c}
\sum_{{\bf k}'}
	{\rm Tr}\, \langle \hat\Xi_{{\bf k}' \xi}^{\vphantom{\dag}} \rangle
=
\frac{1}{\bar \Gamma_1 - \bar \Gamma_2}
\left(
	\bar \Gamma_1 {\rm Tr}\, \hat \Delta_\xi^{\vphantom{\dag}}
	-
	\bar \Gamma_2 {\rm Tr}\, \hat\Delta_\xi^\dag
\right)
+
\frac{{\cal E}_0}{2}
\frac{1}{N_c}
\sum_{{\bf k}' \xi'}
	{\rm Tr}\,
	\langle
		\hat\Xi_{{\bf k}' \xi'}^{\vphantom{\dag}}
		+
		\hat\Xi_{{\bf k}' \xi'}^\dag
	\rangle
-
e \Phi.
\end{eqnarray}
Performing summation over
$\xi$,
we derive
\begin{eqnarray}
%%%%%%%%%%%%%%%%%%%%%%%%%%%%%%%%%%%%%%%%%%%%%%%%%%
\label{Xi_VS_Delta_Tr2}
%%%%%%%%%%%%%%%%%%%%%%%%%%%%%%%%%%%%%%%%%%%%%%%%%%
\frac{(\bar \Gamma_1 + \bar \Gamma_2)}{N_c}
\sum_{{\bf k}' \xi'}
	{\rm Tr}\, \langle
		\hat\Xi_{{\bf k}'\xi'}^{\vphantom{\dag}}
		+
		\hat\Xi_{{\bf k}'\xi'}^\dag
	\rangle
=
\sum_{\xi'}
	{\rm Tr}\, \left(
		\hat \Delta_{\xi'}^{\vphantom{\dag}}
		+
		\hat\Delta_{\xi'}^\dag
	\right)
+
2{ \cal E}_0
\frac{1}{N_c}
\sum_{{\bf k}' \xi'}
	{\rm Tr}\,
	\langle
		\hat\Xi_{{\bf k}' \xi'}^{\vphantom{\dag}}
		+
		\hat\Xi_{{\bf k}' \xi'}^\dag
	\rangle
-
 4 e \Phi.
\end{eqnarray}
Therefore,
\begin{eqnarray}
%%%%%%%%%%%%%%%%%%%%%%%%%%%%%%%%%%%%%%%%%%%%%%%%%%
\label{Xi_VS_Delta_Tr3_prime}
%%%%%%%%%%%%%%%%%%%%%%%%%%%%%%%%%%%%%%%%%%%%%%%%%%
\frac{1}{N_c}
\sum_{{\bf k}' \xi'}
	{\rm Tr}\, \langle
		\hat\Xi_{{\bf k}'\xi'}^{\vphantom{\dag}}
		+
		\hat\Xi_{{\bf k}'\xi'}^\dag
	\rangle
=
\frac{4 e \Phi}{2{\cal E}_0 - \bar \Gamma_1 - \bar \Gamma_2}
-
\frac{1}{2{\cal E}_0 - \bar \Gamma_1 - \bar \Gamma_2}
\sum_{\xi'}
	{\rm Tr}\, \left(
		\hat \Delta_{\xi'}^{\vphantom{\dag}}
		+
		\hat\Delta_{\xi'}^\dag
	\right).
\end{eqnarray}
We can express
$\rho_{10} - \rho_{20}$
in terms of the order parameter as follows
\begin{eqnarray}
%%%%%%%%%%%%%%%%%%%%%%%%%%%%%%%%%%%%%%%%%%%%%%%%%%
\label{n_VS_Delta}
%%%%%%%%%%%%%%%%%%%%%%%%%%%%%%%%%%%%%%%%%%%%%%%%%%
\frac{\rho_{10} - \rho_{20}}{N_c}
=
- \frac{1}{ N_c}
\sum_{{\bf k}' \xi'}
	{\rm Tr}\, \langle
		\hat\Xi_{{\bf k}'\xi'}^{\vphantom{\dag}}
		+
		\hat\Xi_{{\bf k}'\xi'}^\dag
	\rangle
=
-
\frac{4 e \Phi}{2{\cal E}_0 - \bar \Gamma_1 - \bar \Gamma_2}
+
\frac{1}{2{\cal E}_0 - \bar \Gamma_1 - \bar \Gamma_2}
\sum_{\xi'}
	{\rm Tr}\, \left(
		\hat \Delta_{\xi'}^{\vphantom{\dag}}
		+
		\hat\Delta_{\xi'}^\dag
	\right).
\end{eqnarray}
Substituting
expression~(\ref{Xi_VS_Delta_Tr3_prime})
for the trace into the right-hand side of
Eq.~(\ref{Xi_VS_Delta2}),
we establish
\begin{eqnarray}
%%%%%%%%%%%%%%%%%%%%%%%%%%%%%%%%%%%%%%%%%%%%%%%%%%
\label{Xi_VS_Delta_Tr4}
%%%%%%%%%%%%%%%%%%%%%%%%%%%%%%%%%%%%%%%%%%%%%%%%%%
&&\frac{1}{N_c}
\sum_{{\bf k}'}
	\langle \hat\Xi_{{\bf k}' \xi}^{\vphantom{\dag}} \rangle
=
\frac{1}{\bar \Gamma_1^2 - \bar \Gamma_2^2}
\left(
	\bar \Gamma_1 \hat \Delta_\xi^{\vphantom{\dag}}
	-
	\bar \Gamma_2 \hat\Delta_\xi^\dag
\right)
-
\\
\nonumber
&&\left\{
	\frac{{\cal E}_0/4}{
		(\bar \Gamma_1 + \bar \Gamma_2)
		(2{\cal E}_0 - \bar \Gamma_1 - \bar \Gamma_2)
	}
	\sum_{\xi'}
		{\rm Tr}\, \left(
			\hat \Delta_{\xi'}^{\vphantom{\dag}}
			+
			\hat\Delta_{\xi'}^\dag
		\right)
	-
	\frac{2{\cal E}_0 e \Phi}{
		(\bar \Gamma_1 + \bar \Gamma_2)
		(2{\cal E}_0 - \bar \Gamma_1 - \bar \Gamma_2)
	}
\right\}
\hat{\mathbb{I}}_2
-
\frac{e \Phi}{2(\bar \Gamma_1 + \bar \Gamma_2)}
\hat{\mathbb{I}}_2,
\end{eqnarray}
Thus,
\begin{eqnarray}
%%%%%%%%%%%%%%%%%%%%%%%%%%%%%%%%%%%%%%%%%%%%%%%%%%
\label{Xi_VS_Delta_Tr_APP}
%%%%%%%%%%%%%%%%%%%%%%%%%%%%%%%%%%%%%%%%%%%%%%%%%%
\nonumber
\frac{1}{N_c}
\sum_{{\bf k}'}
	\langle \hat\Xi_{{\bf k}' \xi}^{\vphantom{\dag}} \rangle
&=&
\frac{1}{\bar \Gamma_1^2 - \bar \Gamma_2^2}
\left(
	\bar \Gamma_1 \hat \Delta_\xi^{\vphantom{\dag}}
	-
	\bar \Gamma_2 \hat\Delta_\xi^\dag
\right)
-
\frac{{\cal E}_0}{
	4(\bar \Gamma_1 + \bar \Gamma_2)
	(2{\cal E}_0 - \bar \Gamma_1 - \bar \Gamma_2)
}
\sum_{\xi'}
	{\rm Tr}\, \left(
		\hat \Delta_{\xi'}^{\vphantom{\dag}}
		+
		\hat\Delta_{\xi'}^\dag
	\right)
\hat{\mathbb{I}}_2\\
&+&
\frac{e \Phi}
{2({2\cal E}_0 - \bar \Gamma_1 - \bar \Gamma_2)}
\hat{\mathbb{I}}_2.
\end{eqnarray}

We can use Eq.~\eqref{eq:first_self_consistency} and and exclude the term 
$\frac{1}{N_c}\sum_{{\bf k}'} 	\langle \hat\Xi_{{\bf k}' \xi}^{\vphantom{\dag}} \rangle$
from the latter equation.
As a result, we obtain non-linear equations, which can be solved to calculate the order parameter matrices
$\hat{\Delta}_\xi^{\vphantom{\dag}}$
and
$\hat{\Delta}_\xi^\dag$.
If we assume that the order parameter is Hermitian,
$\hat{\Delta}_\xi^{\vphantom{\dag}} = \hat{\Delta}_\xi^\dag$,
then, such equations read as
\begin{eqnarray}
%%%%%%%%%%%%%%%%%%%%%%%%%%%%%%%%%%%%%%%%%%%%%%%%%%
\label{sefl_cons_Hermit}
%%%%%%%%%%%%%%%%%%%%%%%%%%%%%%%%%%%%%%%%%%%%%%%%%%
\frac{1}{2N_c}\sum_{\bf k}\sum_{i=1}^2
    \frac{D_{\xi i}(v_{\xi i} v_{\xi i}^\dagger)}{\sqrt{\varepsilon_{\mathbf k}^2+D_{\xi i}^2}}\Theta\left(\sqrt{\varepsilon_{\mathbf k}^2+D_{\xi i}^2}-\mu\right)
\nonumber\\
=\frac{\hat \Delta_\xi^{\vphantom{\dag}}}{\bar \Gamma_1 + \bar \Gamma_2}
-
\frac{{\cal E}_0}{
	2(\bar \Gamma_1 + \bar \Gamma_2)
	(2{\cal E}_0 - \bar \Gamma_1 - \bar \Gamma_2)
}
\hat{\mathbb{I}}_2
\sum_{\xi'}
	{\rm Tr}\, \hat \Delta_{\xi'}^{\vphantom{\dag}}
+
\frac{e \Phi }
{2(2{\cal E}_0 - \bar \Gamma_1 - \bar \Gamma_2)}
\hat{\mathbb{I}}_2.
\end{eqnarray}
\end{widetext}
As the order parameter is Hermitian, we diagonalize it by a suitable
unitary transformation. Thus, instead of two matrix equations (one per
valley), four coupled scalar equations emerge. After summation over
$\mathbf{k}$
and evident transformations we obtain from
Eq.~\eqref{sefl_cons_Hermit}
the self-consistency
equation~\eqref{eq:self_consistency_dimensionful}.

\section{Numerical procedure for solving the self-consistency equations}
\label{appendix:numerical_procedure}

We employ the following strategy for the numerical solution of the self-consistency equations~\eqref{eq::self_consistency_equation}. 
First, for the given value of $M$ we find special points (points of extremes and inflection) of the function $C(x; M)$ and identify monotonic branches of this function. In so doing, we obtain the ranges of $C$ that correspond to different numbers of branches. 
For such a range, we find all possible sets $\{x_m\}$ of the roots of Eq.~\eqref{eq::self_consistency_equation}. 
For example, in the case of three branches, the number of such sets is $\frac{(4+3-1)!}{4!(3-1)!}=15$,
while for five branches, we have $\frac{(4+5-1)!}{4!(5-1)!}=70$.  Thus, we obtain a number of combinations $\{x_m(C)\}$ that involve numeric roots of Eq.~\eqref{eq:C(x)}. For each combination, the function $V(C)=C+\Lambda\sum_m x_m(C)$ is calculated. 
The extrema of $V(C)$ have been found numerically and all values of $C$ that yield $V(C)$ equal to the given $V$ have been obtained.
A systematic search for all solutions $x_m$ that correspond to a given particle number is the next challenging problem. 
That is why we solved the self-consistency equations at many different values of $M$ and then compared the Helmholtz free energies falling in a small interval between $n$ and $n+\Delta n$. As a result, we can determine the ground state of the model at given $n$ and $V$.

\bibliography{references}

@article{bilayer_review2016,
  author = {Rozhkov, A. and Sboychakov, A. and Rakhmanov, A. and Nori, F.},
  title = {Electronic properties of graphene-based bilayer systems},
  journal = {Phys. Rep.},
  volume = {648},
  pages = {1 },
  year = {2016}
}

@article{Kotov2012RevModPhys,
  author = {Kotov, V. N. and Uchoa, B. and Pereira, V. M. and Guinea, F. and Castro Neto, A. H.},
  title = {Electron-Electron Interactions in Graphene: Current Status and Perspectives},
  journal = {Rev. Mod. Phys.},
  volume = {84},
  pages = {1067},
  year = {2012}
}

@article{Bao2012,
  author = {Bao, W. and Velasco, J. and Zhang, F. and Jing, L. and Standley, B. and Smirnov, D. and Bockrath, M. and MacDonald, A. H. and Lau, C. N.},
  title = {Evidence for a spontaneous gapped state in ultraclean bilayer graphene},
  journal = {PNAS},
  volume = {109},
  pages = {10802},
  year = {2012}
}

@article{Martin2010,
  author = {Martin, J. and Feldman, B. E. and Weitz, R. T. and Allen, M. T. and Yacoby, A.},
  title = {Local Compressibility Measurements of Correlated States in Suspended Bilayer Graphene},
  journal = {Phys. Rev. Lett.},
  volume = {105},
  pages = {256806},
  year = {2010}
}

@article{Weitz2010,
  author = {Weitz, R. T. and Allen, M. T. and Feldman, B. E. and Martin, J. and Yacoby, A.},
  title = {Broken-Symmetry States in Doubly Gated Suspended Bilayer Graphene},
  journal = {Science},
  volume = {330},
  pages = {812},
  year = {2010}
}

@Article{Mayorov2011,
  author  = {Mayorov, A. S. and Elias, D. C. and Mucha-Kruczynski, M. and Gorbachev, R. V. and Tudorovskiy, T. and Zhukov, A. and Morozov, S. V. and Katsnelson, M. I. and Fal'ko, V. I. and Geim, A. K.},
  citationkey = {Mayorov2011},
  journal = {Science},
  pages   = {860},
  title   = {Interaction-Driven Spectrum Reconstruction in Bilayer Graphene},
  volume  = {333},
  year    = {2011},
}

@article{Freitag2012,
  author = {Freitag, F. and Trbovic, J. and Weiss, M. and Schönenberger, C.},
  title = {Spontaneously Gapped Ground State in Suspended Bilayer Graphene},
  journal = {Phys. Rev. Lett.},
  volume = {108},
  pages = {076602},
  year = {2012}
}

@article{Freitag20122053,
  author = {Freitag, F. and Weiss, M. and Maurand, R. and Trbovic, J. and Schönenberger, C.},
  title = {Homogeneity of bilayer graphene},
  journal = {Solid State Communications},
  volume = {152},
  pages = {2053 },
  year = {2012}
}

@article{veligura2012,
  author = {Veligura, A. and  van Elferen, H.J. and Tombros, N. and Maan, J. C. and Zeitler, U. and van Wees, B.J.},
  title = {Transport gap in suspended bilayer graphene at zero magnetic field},
  journal = {Phys. Rev. B},
  volume = {85},
  pages = {155412},
  year = {2012}
}

@article{Velasco2012,
  author = {Velasco Jr., J. and Jing, L. and Bao, W. and Lee, Y. and Kratz, P. and Aji, V. and Bockrath, M. and Lau, C. N. and Varma, C. and Stillwell, R. and Smirnov, D. and Zhang, Fan and Jung, J. and MacDonald, A. H.},
  title = {Transport spectroscopy of symmetry-broken insulating states in bilayer graphene},
  journal = {Nat. Nanotechnol.},
  volume = {7},
  pages = {156},
  year = {2012}
}

@article{freitag2013,
  author = {Freitag, F. and Weiss, M. and Maurand, R. and Trbovic, J. and Schönenberger, C.},
  title = {Spin symmetry of the bilayer graphene ground state},
  journal = {Phys. Rev. B},
  volume = {87},
  pages = {161402},
  year = {2013}
}

@article{trilayer_quarter2021exper,
    author = {Zhou, Haoxin
and Xie, Tian
and Ghazaryan, Areg
and Holder, Tobias
and Ehrets, James R.
and Spanton, Eric M.
and Taniguchi, Takashi
and Watanabe, Kenji
and Berg, Erez
and Serbyn, Maksym
and Young, Andrea F.},
    title = {Half- and quarter-metals in rhombohedral trilayer graphene},
    journal = {Nature},
    volume = {598},
    pages = {429-433},
    year = {2021}
}

@article{Seiler2022,
  author = {Seiler, A. M. and Geisenhof, F. R. and Winterer, F. and Watanabe, K. and Taniguchi, T. and Xu, T. and Zhang, F. and Weitz, R. T.},
  title = {Quantum cascade of correlated phases in trigonally warped bilayer graphene},
  journal = {Nature},
  volume = {608},
  pages = {298},
  year = {2022}
}

@article{zhou2022isospin,
  author = {Zhou, H. and Holleis, L. and Saito, Y. and Cohen, L. and Huynh, W. and Patterson, C. L. and Yang, F. and Taniguchi, T. and Watanabe, K. and Young, A. F.},
  title = {Isospin magnetism and spin-polarized superconductivity in Bernal bilayer graphene},
  journal = {Science},
  volume = {375},
  pages = {774},
  year = {2022}
}

@article{MCCANN2007110,
  author = {McCann, E. and Abergel, D. S. and Fal'ko, V. I.},
  title = {Electrons in bilayer graphene},
  journal = {Solid State Commun.},
  volume = {143},
  pages = {110},
  year = {2007}
}

@article{Nandkishore2010,
  author = {Nandkishore, R. and Levitov, L.},
  title = {Dynamical Screening and Excitonic Instability in Bilayer Graphene},
  journal = {Phys. Rev. Lett.},
  volume = {104},
  pages = {156803},
  year = {2010}
}

@article{Nandkishore2010b,
  author = {Nandkishore, R. and Levitov, L.},
  title = {Quantum anomalous Hall state in bilayer graphene},
  journal = {Phys. Rev. B},
  volume = {82},
  pages = {115124},
  year = {2010}
}

@article{vafek_rg2010,
  author = {Vafek, O.},
  title = {Interacting fermions on the honeycomb bilayer: From weak to strong coupling},
  journal = {Phys. Rev. B},
  volume = {82},
  pages = {205106},
  year = {2010}
}

@article{vafek_nemat_rg2010,
  author = {Vafek, O. and Yang, K.},
  title = {Many-body instability of Coulomb interacting bilayer graphene: Renormalization group approach},
  journal = {Phys. Rev. B},
  volume = {81},
  pages = {041401},
  year = {2010}
}

@Article{Lemonik2010,
  author  = {Lemonik, Y. and Aleiner, I. L. and Toke, C. and Fal'ko, V. I.},
  citationkey = {Lemonik2010},
  journal = {Phys. Rev. B},
  pages   = {201408},
  title   = {Spontaneous symmetry breaking and Lifshitz transition in bilayer graphene},
  volume  = {82},
  year    = {2010},
}

@article{Jung2011,
  author = {Jung, J. and Zhang, F. and MacDonald, A. H.},
  title = {Lattice theory of pseudospin ferromagnetism in bilayer graphene: Competing interaction-induced quantum Hall states},
  journal = {Phys. Rev. B},
  volume = {83},
  pages = {115408},
  year = {2011}
}

@article{cvetkovic_multi2012,
  author = {Cvetkovic, V. and Throckmorton, R. E. and Vafek, O.},
  title = {Electronic multicriticality in bilayer graphene},
  journal = {Phys. Rev. B},
  volume = {86},
  pages = {075467},
  year = {2012}
}

@article{aa_graph_prl2012,
  author = {Rakhmanov, A. L. and Rozhkov, A. V. and Sboychakov, A. O. and Nori, F.},
  title = {Instabilities of the $AA$-Stacked Graphene Bilayer},
  journal = {Phys. Rev. Lett.},
  volume = {109},
  pages = {206801},
  year = {2012}
}

@article{haritonov_afm2012,
  author = {Kharitonov, M.},
  title = {Antiferromagnetic state in bilayer graphene},
  journal = {Phys. Rev. B},
  volume = {86},
  pages = {195435},
  year = {2012}
}

@article{bilayer_half-metal2013numeric_MF,
  author = {Yuan, J. and Xu, H. and Wang, H. and Zhou, Y. and Gao, H. and Zhang, C.},
  title = {Possible half-metallic phase in bilayer graphene: Calculations based on mean-field theory applied to a two-layer Hubbard model},
  journal = {Phys. Rev. B},
  volume = {88},
  pages = {201109},
  year = {2013}
}

@article{baima2018dft_half_met_graphene,
  author = {Baima, J. and Mauri, F. and Calandra, M.},
  title = {Field-effect-driven half-metallic multilayer graphene},
  journal = {Phys. Rev. B},
  volume = {98},
  pages = {075418},
  year = {2018}
}

@article{aa_graph_BreyFertig2013,
  author = {Brey, L. and Fertig, H. A.},
  title = {Gapped phase in {$AA$}-stacked bilayer graphene},
  journal = {Phys. Rev. B},
  volume = {87},
  pages = {115411},
  year = {2013}
}

@article{sboychakov2013AA,
  author = {Sboychakov, A. O. and Rakhmanov, A. L. and Rozhkov, A. V. and Nori, F.},
  title = {Metal-insulator transition and phase separation in doped $AA$-stacked graphene bilayer},
  journal = {Phys. Rev. B},
  volume = {87},
  pages = {121401},
  year = {2013}
}

@article{aa_quarter_met2021,
  author = {Sboychakov, A. O. and Rakhmanov, A. L. and Rozhkov, A. V. and Nori, F.},
  title = {Bilayer graphene can become a fractional metal},
  journal = {Phys. Rev. B},
  volume = {103},
  pages = {L081106},
  year = {2021}
}

@article{Geisenhof2022,
  author = {Geisenhof, F. R. and Winterer, F. and Seiler, A. M. and Lenz, J. and Zhang, F. and Weitz, R. T.},
  title = {Impact of Electric Field Disorder on Broken-Symmetry States in Ultraclean Bilayer Graphene},
  journal = {Nano Lett.},
  volume = {22},
  pages = {7378},
  year = {2022}
}

@article{ab_supercond2023sboychakov,
  author = {Sboychakov, A. O. and Rozhkov, A. V. and Rakhmanov, A. L.},
  title = {Triplet superconductivity and spin density wave in biased AB bilayer graphene},
  journal = {Phys. Rev. B},
  volume = {108},
  pages = {184503},
  year = {2023}
}

@Article{rozhkov2025ab_su4,
  author  = {Rozhkov, A. V. and Sboychakov, A. O. and Rakhmanov, A. L.},
  citationkey = {rozhkov2025ab_su4},
  journal = {Phys. Rev. B},
  pages   = {205421},
  title   = {Ordered states in {AB} bilayer graphene in a {SU(4)}-symmetric model},
  volume  = {111},
  year    = {2025},
}

@Article{rozhkov2023aa_su4,
  author  = {Rozhkov, A. V. and Sboychakov, A. O. and Rakhmanov, A. L.},
  citationkey = {rozhkov2023aa_su4},
  journal = {Phys. Rev. B},
  pages   = {205153},
  title   = {Ordering in the {SU(4)}-symmetric model of {AA} bilayer graphene},
  volume  = {108},
  year    = {2023},
}

@article{rozhkov2026bias_cascade,
  title={Ordered states of undoped AB bilayer graphene: Bias-induced cascade of transitions},
  author={Rozhkov, AV and Sboychakov, AO and Rakhmanov, AL},
  journal={Phys. Rev. B},
  volume={113},
  number={23},
  pages={235421},
  year={2026},
  publisher={APS}
}

@article{mayrhofer2025valley,
  title={Valley-and spin-polarized states in Bernal bilayer graphene},
  author={Mayrhofer, R David and Chubukov, Andrey V},
  journal={Phys. Rev. B},
  volume={111},
  number={24},
  pages={245114},
  year={2025},
  publisher={APS}
}

@article{de_la_barrera_cascade_2022,
    title = {Cascade of isospin phase transitions in {Bernal}-stacked bilayer graphene at zero magnetic field},
    volume = {18},
    copyright = {2022 The Author(s), under exclusive licence to Springer Nature Limited},
    issn = {1745-2481},
    url = {https://www.nature.com/articles/s41567-022-01616-w},
    doi = {10.1038/s41567-022-01616-w},
    number = {7},
    urldate = {2026-02-05},
    journal = {Nature Physics},
    author = {de la Barrera, Sergio C. and Aronson, Samuel and Zheng, Zhiren and Watanabe, Kenji and Taniguchi, Takashi and Ma, Qiong and Jarillo-Herrero, Pablo and Ashoori, Raymond},
    month = jul,
    year = {2022},
    note = {Publisher: Nature Publishing Group},
    pages = {771--775},
}

@Article{Holzmann2020,
  author    = {Holzmann, Markus and Moroni, Saverio},
  citationkey = {Holzmann2020},
  journal   = {Phys. Rev. Lett.},
  pages     = {206404},
  title     = {Itinerant-Electron Magnetism: The Importance of Many-Body Correlations},
  volume    = {124},
  year      = {2020},
  month     = {May},
  doi       = {10.1103/PhysRevLett.124.206404},
  issue     = {20},
  numpages  = {5},
  publisher = {American Physical Society},
  url       = {https://link.aps.org/doi/10.1103/PhysRevLett.124.206404},
}

@Article{Raines2024,
  author    = {Raines, Zachary M. and Glazman, Leonid I. and Chubukov, Andrey V.},
  citationkey = {Raines2024},
  journal   = {Phys. Rev. Lett.},
  pages     = {146501},
  title     = {Unconventional Discontinuous Transitions in Isospin Systems},
  volume    = {133},
  year      = {2024},
  month     = {Oct},
  doi       = {10.1103/PhysRevLett.133.146501},
  issue     = {14},
  numpages  = {6},
  publisher = {American Physical Society},
  url       = {https://link.aps.org/doi/10.1103/PhysRevLett.133.146501},
}
\FloatBarrier

\end{document}